\documentclass[english,aip, numerical,jcp,reprint,twocolumn,floatfix]{revtex4-1}
\usepackage{natbib,hyperref}
\usepackage{amsmath}
\usepackage{graphicx}
\usepackage{hyperref} 
\usepackage{dcolumn}
\usepackage{bm}
\usepackage{epsfig,color,xspace,multirow,xr,bbold}
\usepackage[all]{xy}
\usepackage{setspace}
\usepackage{url}
\usepackage[colorinlistoftodos]{todonotes}
\usepackage{threeparttable} 
\usepackage{natbib,hyperref}
\usepackage{float}
\usepackage{placeins}
\usepackage[utf8]{inputenc}

\newcommand*{\ditto}{\texttt{"}}
\newcommand{\bigqm}{bigQM7$\rm{\omega}$\,}
\newcommand{\cebe}{$E_{\rm b}^{1s}$\,}
\newcommand{\dscf}{$\Delta$-SCF\,}
\newcommand*{\ie}{{\it i.e.,}\,}
\usepackage[normalem]{ulem}
\usepackage{xcolor}
\newcommand{\RRef}[1]{Ref.~\onlinecite{#1}}

\usepackage{soul} 
\begin{document}

\title{
Chemical Space-Informed Machine Learning Models for
Rapid Predictions of X-ray Photoelectron Spectra of Organic Molecules
}
\date{\today}

\author{Susmita Tripathy}
\affiliation{Tata Institute of Fundamental Research, Hyderabad 500046, India}

\author{Surajit Das}
\affiliation{Tata Institute of Fundamental Research, Hyderabad 500046, India}

\author{Shweta Jindal}
\affiliation{Tata Institute of Fundamental Research, Hyderabad 500046, India}

\author{Raghunathan Ramakrishnan}
\email{ramakrishnan@tifrh.res.in}
\affiliation{Tata Institute of Fundamental Research, Hyderabad 500046, India}


\begin{abstract}
We present machine learning models based on kernel-ridge regression for predicting X-ray photoelectron spectra of organic molecules
originating from the $K$-shell ionization energies of carbon (C), nitrogen (N), oxygen (O), and fluorine (F) atoms. We constructed the training dataset through high-throughput 
calculations of $K$-shell core-electron binding energies (CEBEs)
for 12,880 small organic molecules in the \bigqm dataset, employing the $\Delta$-SCF formalism coupled with meta-GGA-DFT  
and a variationally converged basis set. The models are cost-effective, as they require the atomic coordinates of a molecule generated using universal force fields while estimating the target-level CEBEs corresponding to DFT-level equilibrium geometry. 
We explore transfer learning by utilizing the atomic environment feature vectors learned using a graph neural network framework in kernel-ridge regression. Additionally, we enhance accuracy within the $\Delta$-machine learning framework by leveraging inexpensive baseline spectra derived from Kohn--Sham eigenvalues. When applied to 208 combinatorially substituted uracil molecules larger than those in the training set, our analyses suggest that the models may not provide quantitatively accurate predictions of CEBEs but offer a strong linear correlation relevant for 
virtual high-throughput screening.
We present the dataset and models as the Python module, {\tt cebeconf}, to facilitate further explorations.
\end{abstract}

\maketitle
\section{Introduction}

X-ray photoelectron spectroscopy (XPS) distinguishes atoms based on 
the ionization energies of their core electrons
effectively screened by molecular valence electrons. 
The resulting chemical shifts of core-electron binding energies (CEBEs) enable the 
identification of the local bonding environment of an atom. 
A typical XPS spectrum comprises peaks; some merged to form a broader distribution, corresponding to ionized core electrons detected experimentally as photocurrent\cite{groot2008core}. 
The positions and intensities of an XPS spectrum can provide insights on molecular internal coordinates\cite{bagus2022xps}, orientation of adsorbates relative to the substrate\cite{diller2014temperature}, chemical composition\cite{ayiania2020deconvoluting,feng2010gas}, as well as physical conditions such as temperature\cite{diller2014temperature,ayiania2020deconvoluting} and pressure\cite{feng2010gas}.

Advancements in modern synchrotron sources have enabled the precise detection of chemical shifts of CEBEs, achieving a resolution of 0.05 eV \cite{willmott2019introduction}.
The precision of 1$s$-CEBEs (denoted hitherto \cebe) in atoms is limited by lifetime broadening. 
XPS achieves a precision of $0.1$ eV for 1$s$-CEBEs of carbon (C) atoms in molecules\cite{kovavc2014characterisation}.
For nitrogen (N) and oxygen (O), 
peak widths of 0.13 and 0.16 eV, respectively, are commonly used in spectral deconvolution\cite{kovavc2014characterisation}, often corroborated by theoretical peak assignments\cite{ayiania2020deconvoluting}. 
Theoretical CEBEs of even semi-quantitative accuracy can aid 
experimental XPS assignment\cite{azuara2023n, greczynski2020x}, thereby 
contributing to diverse applications such as 
structure elucidation\cite{diller2014temperature}, catalysis\cite{trinh2018synergistic,nguyen2019understanding}, 
characterization of energy storage materials\cite{yu2022degradation}, and
material composition analysis\cite{azuara2023n},

For achieving quantitatively accurate theoretical predictions of XPS, it is crucial to
account for core-orbital relaxation effects, which unfold at a sub-femtosecond ($<10^{-15}$ s) timescale mirroring the ultrafast temporal dynamics during experimental detection, resonating with the energy-time uncertainty principle\cite{groot2008core, kohiki1999many}. CEBEs estimated using molecular orbital (MO) energies---based on the Koopmans' approximation\cite{chong2002interpretation, azuara2023n} in
density functional theory (DFT)---show large errors due to the lack of 
core relaxation in MOs that determine the ground state electronic density\cite{bagus2013interpretation, besley2021modeling}. 
Many-body methods such as $GW$ approximation\cite{aryasetiawan1998gw}, which utilizes Green's function ($G$) and a screened Coulomb interaction ($W$), and random phase approximation\cite{ren2012random} adequately address screening effects. While these
methods provide robust frameworks, their computational complexity prohibits their application for large-scale data generation.  
Within DFT, alternative approaches such as the Slater transition state method \cite{williams1975generalization, jana2023slater} and the $\Delta$-self consistent field (\dscf) approach \cite{bagus1965self, kahk2019accurate, bellafont2016performance} offer promising avenues for incorporating orbital relaxation effects. 
These methods explicitly account for the vacancy in a 
core-orbital---{\it i.e.}, hole---using charge-constraining algorithms such as the maximum-overlap-method \cite{gilbert2008self}, thereby preventing the variational collapse of non-Aufbau Slater determinants. While the numerical convergence \cite{carter2020state} of \dscf is highly coupled to the choice of projectors used for wavefunction localization \cite{klein2021nuts, behler2007nonadiabatic}, it emerges as the best option in terms of the overall efficiency\cite{michelitsch2019efficient, klein2021nuts}.

\begin{figure*}[htbp]
        \centering
        \includegraphics[width=0.8\linewidth]{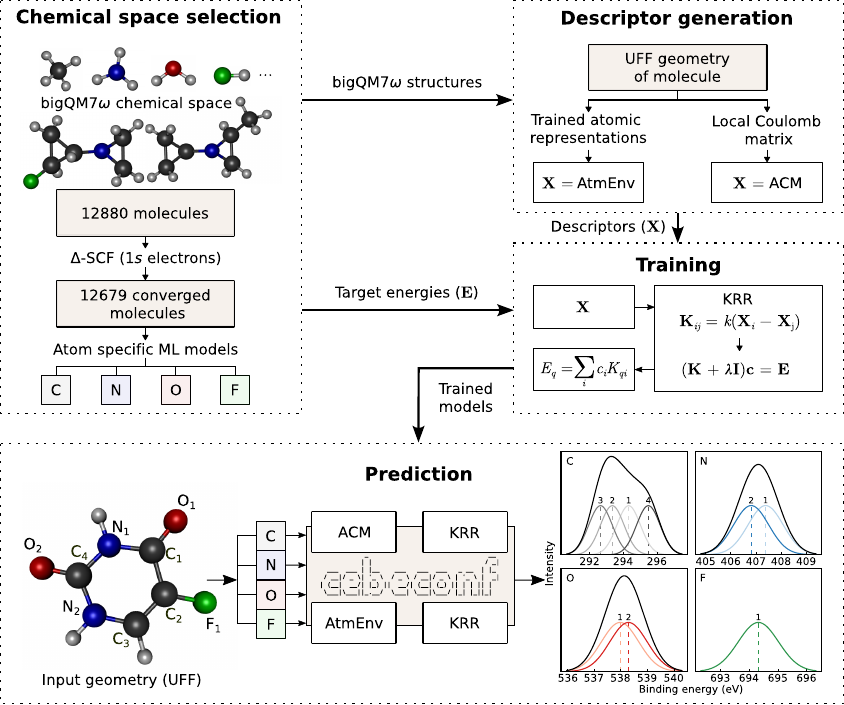}
        \caption{
        Workflow of the study: 
        (i) For 85,837 C, O, N, or F atoms in 12,679
        molecules of the \bigqm chemical space, \dscf calculations were performed to
        calculate 1$s$-CEBEs.
        (ii) Structure-based descriptors---atomic environment from graph neural networks (AtmEnv) and atomic Coulomb matrix (ACM)---were generated using
        molecular geometries calculated with UFF. 
        (iii) For each main-group element, kernel-ridge-regression models were created using 80\% of the data. 
        (iv) Trained models are shared through the {\tt cebebconf} module\cite{cebeconf}  that requires UFF-level molecular geometry as the input for new predictions.
        }
        \label{fig:workflow}
\end{figure*}

The magnitude of \cebe of an element in its reduced state is lower than its oxidized states. 
For example, the \cebe (C) of a CH moiety is approximately 285~eV\cite{dorey2018xps}, which, with the introduction of electronegative substituents as in CF, CF$_2$, CF$_3$, and CF$_3$OH 
is systematically shifted 
to 288, 291, 293.5, and 295 eV, respectively\cite{willmott2019introduction}.
XPS measurements of fluorinated polyethylenes reveal similar trends\cite{ferraria2003xps}. 
These systematic variations in \cebe with chemical composition reveal a clear correlation between this quasi-atomic property and
the local environment of an atom in a molecule. The formal existence of this structural-property relationship is a crucial prerequisite for ML modeling based on structure-based descriptors, as it ensures that the models can reliably leverage structural information to predict XPS spectra.

Molecular chemical space datasets, such as 
QM9 \cite{ramakrishnan2014quantum} and 
\bigqm\cite{kayastha2022resolution} have facilitated the elucidation of structure-property relationships through 
machine learning (ML) modeling\cite{rupp2015machine}. 
The $\Delta$-ML\cite{ramakrishnan2015big} approach further enhances the accuracy by utilizing inexpensive baseline methods\cite{gupta2021revving, watson2023delta, golze2022accurate}. In recent years, ML models have successfully 
predicted local properties of atoms-in-molecules (AIM) such as nuclear magnetic resonance (NMR) 
shielding parameters\cite{gupta2021revving,shiota2024universal,el2024transfer, el2024global} and acid dissociation constants (pKa)\cite{el2024transfer, el2024global}, 
as well as `global' molecular properties such as 
atomization energies\cite{ramakrishnan2014quantum}, 
electronic excitation energies in ultraviolet (UV)-visible spectroscopy\cite{ramakrishnan2015electronic, gupta2021data, kayastha2022resolution},  or frontier MO energy gaps \cite{fediai2023interpretable}, demonstrating the effectiveness 
of ML-aided computational chemistry endeavors \cite{ramakrishnan2017machine}.
ML techniques have found utility in spectroscopic applications for mapping spectroscopic features to 
structural motifs\cite{kotobi2023integrating,choudhury2024structure}. 
Specifically, several studies \cite{golze2022accurate, aarva2019understanding, zarrouk2024experiment} have explored the prediction of XPS spectra in materials. 
In the context of XPS spectra for gas-phase molecules, Rupp \textit{et al.}\cite{rupp2015machine} investigated a subset of QM9 with the stoichiometry C$_7$H$_{10}$O$_2$, consisting of 6095 constitutional isomers, while Golze \textit{et al.}\cite{golze2022accurate} focused on a QM9 subset containing only H, C, and O atoms. Although the inclusion of additional elements in the training data can reduce the accuracy of ML models, it also enhances their generality. Therefore, we include all CONF atoms from the \bigqm dataset, incorporating greater diversity into our training data. 
The present study explores data-driven modeling of XPS spectra of 
small organic molecules. 
We select the new chemical space dataset, \bigqm\cite{kayastha2022resolution},
which has a larger number of molecules containing 
upto 7 CONF atoms than the corresponding subset of QM9, 
offering greater structural and compositional diversity.
The minimum energy structures in \bigqm were calculated using the 
range-separated hybrid DFT method $\omega$B97XD\cite{chai2008long}, thereby enhancing the quality of structures suitable for developing
datasets of various properties in a single-point fashion. 
We present ML models for predicting the XPS spectra of organic molecules at the \dscf level, trained using structural descriptors derived from inexpensive geometries obtained with the universal force field (UFF) \cite{rappe1992uff}. Previous studies have explored the use of cost-effective baseline-level geometries for out-of-sample predictions without a significant loss in accuracy\cite{ramakrishnan2015big,ramakrishnan2017machine}.
By using UFF-level geometries, we eliminate the need for computationally expensive DFT-level geometries, making the XPS calculations for new ML predictions more efficient.
These inexpensive inputs are used to generate light-weight 
descriptors obtained from graph neural networks (GNNs) and 
a local version of the Coulomb matrix (CM) descriptor\cite{rupp2012fast,rupp2015machine}.
Further, we propose using Kohn--Sham (KS) eigenvalues obtained from single-point calculations of neutral molecules as a baseline in $\Delta$-ML. FIG.~\ref{fig:workflow} illustrates our data generation and ML workflow. The trained models are packaged within the Python module \texttt{cebeconf}\cite{cebeconf}, providing easy access for XPS predictions.

In the following, \ref{sec_elec} compiles details of electronic structure calculations. 
In \ref{sec_mulliken}, we discuss a scheme to assign quasi-particle energies to CEBEs based
on the Mulliken population scheme. 
We describe the details of ML for modeling local atomic properties in \ref{sec_KRR}. 
In \ref{sec_Local}, we gather the details of atomic descriptors. We present our results in \ref{sec_results}: 
Evaluation of the DFT method used for data generation (\ref{sec_method}); 
property trends and data distribution (\ref{sec_trends}); and analysis of the 
performance of ML models (\ref{sec_eval} and \ref{sec_learning}).
We highlight the transferability of our models by applying them to 
a class of biomolecules in \ref{sec_validation}. Finally, we conclude in section \ref{sec_conclusion}, 
highlighting the features of the ML models presented in this study and their scope for future applications.

\section{Methodology}\label{sec_method}

\subsection{Electronic structure calculations}\label{sec_elec}
We performed electronic structure calculations using 
the all-electron, numerically tabulated atom-centered 
orbital (NAO) code, FHI-aims\cite{blum2009ab}. 
For training the ML models reported in this study, we selected
the \bigqm dataset, which has been used for generating datasets of electronic excitation spectra\cite{kayastha2022resolution,majumdar2024resilience}.
The molecular structures in \bigqm were optimized in the original study\cite{kayastha2022resolution}
using the ConnGO approach that preserves the covalent bond connectivities during geometry optimization \cite{senthil2021troubleshooting}. For this purpose, the 
range-separated hybrid DFT method, $\omega$B97XD, was used in combination
with the def2TZVP basis set. 
Since the change in nuclear coordinates have
negligible effects on XPS, the minimum energy geometries of
neutral ground states were used in all calculations.

For \dscf calculations, we selected the 
meta-generalized gradient approximation (mGGA) to DFT, SCAN,
along with the Tight-Full, basis set.  
The 1$s$-CEBE (\cebe) in a \dscf calculation is obtained 
by subtracting the total energy of the neutral molecule ($E_{\rm mol}$) 
from that of the corresponding
1$s$ core-ionized cation ($E_{\rm cation}^{1s}$)
\begin{equation}
    E_{\rm b}^{1s} = E_{\rm cation}^{1s} - E_{\rm mol}. 
    \label{eq_1sCEBE}
\end{equation}
We exclusively use NAOs with `tight' integration grids to accurately account for orbital relaxation effects 
in the cationic states involved in the \dscf calculations.
While `tight' settings in FHI-aims have `tier 2' basis functions as default, 
we selected the highest `tier' possible for each element, \ie `tier 3' for hydrogen (H) 
and beyond `tier 4' for CONF atoms. We refer to this basis set as Tight-Full. We have also probed the 
valence-correlation consistent NAO, NAO-VCC-5Z, which is analogous to 
Dunning's cc-pV5Z\cite{zhang2013numeric}. Since there is minimal error 
for core-levels when using large basis sets, 
core augmentation functions \cite{sarangi2020basis, jana2023slater} were not utilized. 

Holes at core-levels were constrained using the {\tt force\_occupation\_basis} keyword in FHI-aims. The corresponding approach is a variant of the maximum overlap method \cite{gilbert2008self}, specifying the 
hole in a quasi-AO\cite{mulliken1955electronic}. 
One can also use the {\tt force\_occupation\_projector}, which directly constrains the hole 
at a specified MO; however, we found the former to be more variationally stable\cite{klein2021nuts} 
making it suitable for automated high-throughput calculations 
with minimal manual data curation. 

Intricate modifications specific to core-level predictions are avoided in the 
Koopmans' approximation. Accordingly, the negative of the KS orbital energy, $\varepsilon_i$, 
approximates the 1$s$-CEBE\cite{chong2002interpretation, klein2021nuts}
\begin{equation}
    E_{\rm b}^{1s} \approx -\varepsilon_{1s}.
    \label{eq_koopmans}
\end{equation}
In this work, we performed single point calculations 
with the PBE-DFT method and the cc-pVDZ basis set to 
calculate  $\varepsilon_{1s}$ using
UFF-level structures obtained
using the program Openbabel\cite{o2011open}.
Despite lacking contributions from orbital relaxation effects in the Koopmans' interpretation \cite{bellafont2015prediction, bagus2013interpretation}, its relative ease of use makes 
it an efficient choice as a baseline method in $\Delta$-ML.

$GW$ method is considered 
the ``gold standard'' for modeling CEBEs\cite{golze2022accurate, li2022benchmark}. 
In particular, the contour deformation to evaluate self energy in $GW$, improves the quasi-particle energies for core-levels\cite{golze2018core, golze2019gw}. 
Among presently available $GW$ approximations, Hedin shifted $GW$, $G_{\Delta H}W_0$, which introduces a state-specific global shift in the quasiparticle self-consistent equations 
is the most accurate\cite{li2022benchmark}. 
Hence, in this study we use $G_{\Delta H}W_0$ for evaluating the accuracy of \dscf results. 
In all $GW$ calculations, we used KS--eigenstates obtained using PBE-DFT with cc-pVnZ (n=T,Q,5) basis sets and extrapolated the results
to arrive at the basis limit values, following the procedure previously used by Golze {\it et al.}\cite{golze2019gw, golze2020accurate}. 
For systems exhibiting non-monotonic convergence with basis set size, due to contraction errors\cite{mejia2022basis}, 
we used cc-pV5Z values instead of extrapolated energies. 
The computational complexity of $GW$ methods, which increases manyfold with basis set size limits its applicability in
high-throughput data generation. 
Therefore, $GW$ calculations were performed only for a small subset of \bigqm.
We use $GW$-predicted \cebe for assessing the accuracy \dscf and Koopmans' 
approximations based on the mGGA-DFT methods, SCAN and TPSS.

In our calculations, 219 1$s$-core-ionized cations did not achieve density convergence during the SCF procedure. These atoms belong to 201 molecules, which have been excluded from the dataset of 12,880 molecules. An analysis of these 201 molecules indicates that all of them that experienced convergence failure contain at least one double bond. While the corresponding atoms are not necessarily $sp^2$ hybridized in every instance, double bonds generally contribute to convergence difficulties and are the most common issue among each atom type when classified by hybridization. The remaining molecules exhibiting convergence failure contain a carbonyl group. The distribution of these non-converged systems is detailed in Table S5 of the SI. SMILES entries for all molecules exhibiting SCF convergence failure are listed in Table S6 of the SI. While such calculations can be converged through careful assessments of 
numerical criteria and localization procedures\cite{michelitsch2019efficient}, 
we excluded these molecules for the sake
of consistent numerical settings across the dataset. 

The resulting XPS dataset consists of 85837 entries of \cebe 
of the constituent CONF atoms of 12679 molecules obtained 
using \dscf with the SCAN/Tight-Full method. 
From this dataset, shuffled train-test splits of various sizes were selected for training the ML models.
Further, to validate the ML models, we explore combinatorially substituted derivatives of 
pyrimidinones and uracil, resulting in 208 biomolecules. 
For this set, we have calculated \cebe at the SCAN/Tight-Full level 
using minimum energy structures determined using the $\omega$B97XD/def2TZVP method
as implemented in Gaussian16\cite{frisch2016gaussian}. 
Further, for the validation set, we have calculated Koopmans' estimations
with the PBE/cc-pVDZ method using geometries determined with UFF.
In all calculations, we accounted for relativistic effects through the 
atomic zeroth-order regular approximation 
(aZORA)\cite{lenthe1996zora, klein2021nuts}. 
When using the $\Delta$-ML models presented in this study for new predictions, along with UFF geometry, it is necessary to provide the baseline CEBEs, which are the PBE/cc-pVDZ Koopmans' estimates determined using UFF geometry. Since the effect of aZORA on CEBEs results in a constant shift, non-relativistic PBE/cc-pVDZ values can be corrected \textit{a posteriori} \cite{bellafont2016performance, golze2020accurate, keller2020relativistic, li2022benchmark} by adding 0.29, 0.55, 0.97, or 1.57 eV for C, O, N, or F atoms, respectively. Further details are provided in Figure~S1 of the SI.

\subsection{Assignment of quasi-particle energy to CEBE}\label{sec_mulliken}

The localization of core-holes is assumed inherently in XPS. However, 
vibronic fine structure underlying the 
XPS spectrum sheds insights on the participation of 
core-level MOs in very-weak bonding\cite{kempgens1997core, myrseth2002vibrational}. 
These splittings can be of the order of milli-eVs and observed in 
high-precision spectroscopic techniques. 
Dispersions in CEBEs for chemically equivalent atoms in symmetric molecules such as benzene\cite{myrseth2002vibrational}, acetylene\cite{kempgens1997core}, 
dinitrogen\cite{hergenhahn2001symmetry}, and others \cite{matz2023ab} have been observed. 
Splittings arising from the coupling of core-levels can also be observed and identified in the one-particle energies obtained from KS-eigenvalues. Experimentally, in the case of benzene, 6 degenerate core levels lead to 4 energy levels spanning a range of 64 meV \cite{myrseth2002vibrational}(see Table~S1 of SI). 
While by definition, the values of \cebe of identical atoms will be similar 
in \dscf predictions, dispersions in \cebe are captured in quasiparticle energies.
Therefore, to map the quasiparticle energies of MOs to the respective \dscf energies for use as a baseline in $\Delta$-ML, we use Mulliken population analysis \cite{mulliken1955electronic}.

For assigning the $k$-th MO to an AO centered on atom $A$,
of angular momentum quantum number $l$, we use 
Mulliken population projected on the MO defined as
\begin{eqnarray}
    P_{A,k,l}^{\rm M} & = & 2 \sum_{\mu \in A, \mu \in l} \sum_{\lambda}  C_{\mu, k} C_{\lambda, k} S_{\lambda,\mu}.
\end{eqnarray}
Here, the summation is performed over all AOs, $\mu$, centered on atom $A$ 
with angular momentum $l$. Here, $C_{\mu, k}$ and $C_{\lambda, k}$ are elements of the 
$k$-th KS eigenvector corresponding to the AOs, $\mu$ and $\lambda$, respectively.
Further, $S_{\lambda,\mu}$ is an element of the overlap or inner product matrix
in the AO representation. 
For a given $k$, the CEBE corresponds to the atom with a maximal projected population. 
Consequently, when the population is identical for symmetrically equivalent atoms, they will be assigned the same quasiparticle energy. 
In the \bigqm dataset, using KS eigenvalues at the PBE/cc-pVDZ level, 79 pairs of chemically equivalent atoms were assigned the same KS-eigenvalue in 59 molecules. 

\subsection{KRR-ML framework for modeling CEBE}\label{sec_KRR}
KRR is an efficient ML framework for accurately modeling molecular and materials properties\cite{hansen2015machine,stuke2019chemical,rupp2012fast}. 
It also facilitates modeling local quasi-atomic properties in molecules such as NMR shielding, CEBE, and partial charges on atoms\cite{rupp2015machine,gupta2021revving}. 
In this study, we apply KRR-ML for modeling \cebe. 
 An attractive feature of KRR is that the training, \ie the model optimization, 
is performed by minimizing a convex function for which direct linear solvers such as 
Cholesky decomposition or matrix inversion yield numerically
exact solution within the limits of numerical precision\cite{scholkopf2002learning, ramakrishnan2017machine}. 
Accordingly, the regression coefficients are obtained by 
solving the following equation.
\begin{eqnarray}
\left[ \textbf{K} + \lambda\textbf{I} \right] \textbf{c} = \textbf{E}_{\rm b}^{1s},     
\label{eq_krr_train}
\end{eqnarray}
where
 $\textbf{K}$ and $\textbf{I}$ are $N\times N$ kernel and identity matrices, respectively, while
 $\textbf{E}_{\rm b}^{1s}$ is the vector of CEBE values of $N$ training entities. 
  The kernel matrix element, $K_{ij}$, captures the 
  similarity between two training atoms ($i$ and $j$) in the form of a kernel function, $k$, 
  sometimes referred as radial basis functions (RBFs)
  of the corresponding descriptors,  $k({\bf d}_i, {\bf d}_j)=k({\bf d}_i - {\bf d}_j)$. 
  For brevity, we define ${\bf d}_i - {\bf d}_j$ as $D_{ij}$.
  The choice of the descriptor or the representation vector, ${\bf d}$, is 
presented in \ref{sec_Local}.
  
 In this study, we explore 
 Laplacian (a.k.a. exponential) and Gaussian kernels defined as 
 \begin{eqnarray}\label{eq_kernels}
k^{\rm Laplacian}(D_{ij})&  = &\exp\left(-||D_{ij}||_1/\sigma\right)\,
{\rm and} \nonumber \\
k^{\rm Gaussian}(D_{ij}) & = & \exp \left(-||D_{ij}||_2^2/ 2 \sigma^2 \right).
\end{eqnarray}
The Laplacian kernel depends on the $L_1$ norm, a.k.a. taxicab norm defined as
$||{\bf x}||_1=\sum_i |x_i|$, whereas 
the Gaussian kernel uses the Euclidean norm
defined as $||{\bf x}||_2=\sqrt{ \sum_i x_i^2}$. 
 There are other definitions of kernels where the pairwise comparison can also be an inner-product\cite{scholkopf2002learning}, which we do not explore in this study. 

For a query atom, $q$,  KRR-ML predicts 
\cebe according to the following equation.
\begin{equation}
     E_{\rm b}^{1s} (q) \approx \sum_{i=1}^N c_i K_{qi},
     \label{eq_krr_pred}
\end{equation}
where the summation on the right side is performed over $N$ training examples, 
c$_i$ are the regression coefficients obtained through training and 
$K_{qi}$ is the kernel matrix element evaluated between the query atom $q$ and training atom $i$.

The hyperparameters, $\lambda$ (in Eq.~\ref{eq_krr_train})
and $\sigma$ (in Eq.~\ref{eq_kernels}), modulate the performance of KRR-ML models. 
 The parameter $\lambda$ is a non-negative real number serving two purposes\cite{ramakrishnan2017machine}. 
For positive values, it introduces a penalty to regularize the magnitudes of the regression coefficients,
 which is necessary to decrease the impact of outliers on the model's performance. 
Another more common scenario is that there are redundancies in the training set.
For instance, training examples $i$ and $j$ are the same resulting in two identical rows and columns in 
 of $\textbf{K}$, which becomes a singular matrix. In such cases, adding the second term, $\lambda{\bf I}$, conditions 
 the linear system defined in Eq.~\ref{eq_krr_train}. 
 It is important to note that the aforementioned redundancies can 
 occur if the representations lack uniqueness, even for a few 
 training examples. In anticipation
 of such linear dependencies, 
 in all KRR-ML calculations, we set $\lambda$ as $10^{-4}$.

The kernel width, $\sigma$, determines the width of the kernel-RBFs, thereby governing the spread over training examples for
predictions.  
In a previous work\cite{ramakrishnan2015many}, a heuristic approach was proposed to determine $\sigma$ by setting the
minimum of the off-diagonal elements of ${\bf K}$ to 1/2, $K_{ij}^{\rm min}=1/2$,
as a measure of conditioning ${\bf K}$. 
Accordingly, $\sigma$ can be determined using the maximal value
of $D_{ij}$ over a random sample. For Laplacian and Gaussian kernels,  one arrives at\cite{ramakrishnan2015many,ramakrishnan2017machine}: 
\begin{equation}\label{eq:sigma_opt_max}
\sigma_{\rm opt}^{\rm Laplacian} = \frac{D_{ij}^{\rm max}}{ \log 2}; \quad
\sigma_{\rm opt}^{\rm Gaussian} = \frac{D_{ij}^{\rm max}}{\sqrt{2 \log 2}}.
\end{equation} 
Since descriptor differences have non-trivial distributions, for 
KRR-ML modeling of NMR shielding in the QM9-NMR dataset \cite{gupta2021revving},
better results were reported when using the median of the descriptor differences instead of
$D_{ij}^{\rm max}$.

To shed more light on the optimal value of kernel-width for a given target property, we set $K_{ij}^{\rm min}$ 
to be free parameter, $0 < \tau < 1$, and determine $\sigma$ as follows:
\begin{equation}\label{eq:sigma_opt_tau}
\sigma_{\rm opt}^{\rm Laplacian} = \frac{D_{ij}^{\rm max}}{ \log 1/\tau}; \quad
\sigma_{\rm opt}^{\rm Gaussian} = \frac{D_{ij}^{\rm max}}{\sqrt{2 \log 1/\tau}}.
\end{equation} 
We scanned $\tau$ in the range 0.03 to 0.99 in steps of 0.03. 

\subsection{Local descriptors for atom-in-molecules}\label{sec_Local}
ML-based modeling of local properties of atoms-in-molecules such as CEBEs presents distinct challenges compared to global molecular or material properties. One of the primary difficulties lies in the lack of specialized structural descriptors tailored for these local properties. Traditional descriptors are designed to capture the overall molecular geometry and are unsuitable for modeling local atomic environments. Additionally, local properties are highly sensitive to subtle changes in the chemical environment, necessitating models that can capture intricate variations at the atomic level. Consequently, more extensive benchmarking of ML strategies is essential to improve the modeling of CEBEs.
For ML modeling of local properties with KRR, several descriptors or 
representations have been proposed to describe the 
atomic environment:  
Gaussian RBFs\cite{behler2007generalized}, 
local CM\cite{rupp2012fast},
smooth overlap of atomic positions (SOAP) \cite{szlachta2014accuracy},
neural network (NN) embeddings\cite{unke2018reactive},
atomic spectral London-Axilrod-Teller-Muto (aSLATM) \cite{huang2020quantum}, and
Faber-Christensen-Huang-Lilienfeld (FCHL)\cite{faber2018alchemical}. 
Formally, many-body RBFs, 
SOAP, aSLATM, and FCHL are continuous representations placing heavy storage
requirements. 
On the other hand, 
CM and embeddings are discrete representations amenable to rapid predictions. 
Since one of the aims of this study is to present ML
models pre-trained on a large training set, 
we have explored atomic CM (ACM) and 
atomic environment obtained as embeddings by 
training a GNN (AtmEnv).

As stated before, of the 12,880 molecules in the 
\bigqm dataset, 201 did not converge during
\dscf calculations. The remaining 12,679 
 were partitioned into four subsets based on the 
 presence of C/O/N/F atoms. 
 The subset with `C' was the largest, with 12674 molecules, 
 as `C' is present in all molecules in \bigqm but HF, H$_2$O, NH$_3$, F$_2$, and O$_2$. 
 See Table~S2 of SI for further details. 

The target property, \cebe of C/O/N/F atoms were 
calculated using the minimum energy geometries of the 
\bigqm molecules, previously determined at the $\omega$B97XD/def2TZVP 
level of theory\cite{kayastha2022resolution}. 
However, to generate local descriptions,
we utilize molecular geometries determined using UFF starting from `simplified molecular-input line-entry system' (SMILES) strings. As in previous studies\cite{ramakrishnan2015big,gupta2021data}, 
 ML models trained using baseline levels such as UFF also 
 capture the change in the structure-property mapping: 
DFT-property@DFT-geometry $\rightarrow$ DFT-property@UFF-geometry.
Training with UFF geometries circumvents computational bottlenecks 
in DFT-level structure optimization, for
out-of-sample queries that are more expensive than
$\Delta$-SCF calculations of \cebe 
enabling rapid application of the ML models to new
systems.

\subsubsection{Atomic CM}
CM is one of the simplest structure-based descriptors 
for mapping atomic coordinates and molecular stoichiometries to a molecular
global property such as atomization energy\cite{rupp2012fast}.
For a molecule with $N$ atoms, CM is an 
$N \times N$ matrix defined as 
\begin{eqnarray}
    M_{ij} & = & Z_{i}^{2.4}/2 ;\quad  i = j \nonumber \\
        & = & Z_{i}Z_{j}/R_{ij}; \quad i \neq j,
    \label{eq:cm}
\end{eqnarray}
where $Z_{i}$ is the atomic number of atom-$i$ while 
$R_{ij}$ is the Euclidean distance between nuclei $i$ and $j$ in \AA.
Changing the unit of coordinates will 
reflect in the kernel-width, $\sigma$.
 The off-diagonal elements in the CM represent Coulomb interaction
 between the nuclei of atoms; the diagonal element 
 is an estimation of the atomic total energy\cite{ramakrishnan2017machine}.

The CM is symmetric and invariant to rotation and translation but 
not to the choice of atomic indices. To make CM atom-index invariant, 
one can uniquely permute its rows and columns based on a metric\cite{von2013first}. The row-sorted norm version is a popular approach\cite{montavon2012learning}. The randomly sorted CM 
approach\cite{montavon2012learning} considers different shuffles, included as 
separate training examples, offering lower prediction errors than the row sorted approach\cite{hansen2013assessment}. As CM is symmetric, either the upper or the lower triangular matrix is sufficient to construct the descriptor vector.

In the atomic CM (ACM), each query atom, $q$, is represented by
an $N \times N$ matrix. 
The indices of the $N$ atoms
are permuted in ascending order of distances of the neighboring atoms
from the query atom, $q$. 
We do not apply a cut-off radius to determine the neighbor atoms. 
To ensure that all the entries have the same descriptor size, the size of ACM
is allocated to $M \times M$ where $M=23$ is the maximum number
of atoms in the \bigqm dataset corresponding to $n$-heptane. All elements of 
the ACM other than the $N \times N$ block are set to zero. 
The indexing of the atoms for ACM is schematically shown in
Figure~S2 of SI.

\subsubsection{AtmEnv: Descriptor from Graph Neural Network}
Inspired by previous research works\cite{el2024global,fediai2023interpretable,jindal2022capturing}, we have selected SchNetpack architecture (version 0.3)\cite{schutt2018schnetpack} to train a descriptor encoding the atomic environment on-the-fly\cite{cho2019enhanced,te2018rgcnn}. 
Some recent works\cite{el2024global,fediai2023interpretable} have demonstrated the use of atomic features from trained GNN models to predict the atomic properties such as pKa, NMR shielding, and frontier MO energies for the QM9 dataset using different ML algorithms. 
The SchNet framework was originally developed to explore molecular potential energy surfaces, making total energy or atomization energy a suitable target for generating embeddings with the GNN. These embeddings are effective for modeling atomic forces due to the direct correlation between total energy and atomic forces. However, to model CEBEs that do not directly correlate with total energy, we have adapted the approach by training SchNet in an unsupervised manner using a null target. The descriptors so obtained from SchNetpack are referred to as `AtmEnv'. After unsupervised training with a null target, the SchNet model, with its optimized weights and embeddings, can be used to extract the AtmEnv vector for a query atom, employing the UFF-level molecular geometry as input. This approach enables the embeddings to be applicable for tasks beyond those directly related to total energy or forces.
Using a null target (i.e., setting target values to zero) allows the embeddings to capture geometric features independent of any specific property, similar to other descriptors like SOAP, ACM, or FCHL\cite{mo2022simple,hamilton2017representation}. 

SchNetpack belongs to the family of convolutional NN and initializes the atomic descriptors as a basis set expansion in atomic numbers, $Z_i$
 \begin{equation}\label{eqn:input}
    y_i^{(0)} = y_i(Z_i),
\end{equation}
 where $y_i^{(0)}$ is the initial feature vector (layer = 0) 
 for an atom, $i$, and $y_i(Z_i)$ are the coefficients expanded in nuclear charges, $Z_i$.
 The SchNet architecture consists of atomwise layers and convolution layers. 
 In the atomwise layers the feature vector, $y_i$, for an atom $i$ is updated as 
   \begin{equation}\label{atomwise}
     y_i^{l+1} = W^ly_i^l + b^l
 \end{equation}
 while in the convolution layer, $y_i$ is updated as
 \begin{equation}\label{conv}
     y_i^{l+1} = \sum_j y_j^l \circ W^l(r_{ij}),
 \end{equation}
 where the summation is over all the neighboring atoms of $j$, and the convolution operator, $\circ$, denotes element-wise multiplication.
 In both cases, $y_i^{l+1}$ is the updated feature vector 
 in layer $l$, $W^l$ and $b^l$ are the network weights and bias weights, respectively, for layer, $l$.
 Further, $W^l$ is the filter containing information about all the interatomic distances, $r_{ij}$, and $y_j^l$ are the features of the neighbor atoms in layer, $l$. 
  The convolutional layers in SchNet form the interaction blocks 
where the feature vector of one atom {\it interacts}
with the feature vectors of other neighbors; 
numerous iterations lead to an updated embedding vector.

 The activation function in the network is kept as shifted softplus, 
 defined as $ssp(y) = \log\left[ \exp( y )/2 + 1/2 \right]$. 
The interatomic distances are provided in the form of Gaussian RBFs
 in the interaction blocks leading to learning of the AtmEnv. 
  We used a cut-off radius of 6\AA~to define the neighbor atoms 
  along with 40 Gaussian functions and 4 interaction blocks.
  Further, we limit the length of the AtmEnv vectors to 128, 
  beyond which the accuracy of the KKR-ML models does not 
  improve substantially, see Table~S3.

\section{Results and Discussions}\label{sec_results}

\subsection{Assessment of DFT methods for data-generation} \label{sec_method}
\begin{table}[ht!]
\centering
\caption{Accuracy of DFT-predicted \cebe 
of 32 molecules in \bigqm, each containing
up to three CONF atoms. 
DFT results were derived using $\Delta$-SCF and Koopmans'  
approximations, using reference values obtained from CBS-extrapolated $G_{\Delta H}W_0$@PBE calculations. 
Tight, and VCC-5Z refer to Tight-Full and NAO-VCC-5Z basis sets.
The assessment includes various
error metrics: 
mean signed error, MSE (in eV) calculated as $GW$-DFT,
mean absolute error, MAE (in eV), and
standard deviation of the error, SD (in eV).}

\small\addtolength{\tabcolsep}{1.2pt}
\begin{tabular}[t]{lll rrr}
\hline
\hline 
\multicolumn{1}{l}{Atom (\#)} & 
\multicolumn{1}{l}{Method} & 
\multicolumn{1}{l}{DFT/Basis set} & 
\multicolumn{3}{l}{Error metrics}  \\
\cline{4-6}
\multicolumn{1}{l}{} &
\multicolumn{1}{l}{} &
\multicolumn{1}{l}{} &  
\multicolumn{1}{l}{MSE} &
\multicolumn{1}{l}{MAE} & 
\multicolumn{1}{l}{SD} \\
\hline
C $(\#49)$ & $\Delta$-SCF          & SCAN/Tight    & -0.13 &  0.13 &  0.11 \\
\ditto     & \ditto                & SCAN/VCC-5Z   & -0.08 &  0.09 &  0.11 \\
\ditto     & \ditto                & TPSS/Tight    & -0.06 &  0.08 &  0.09 \\
\ditto     & \ditto                & TPSS/VCC-5Z   & -0.01 &  0.07 &  0.09 \\
\ditto     & $-\varepsilon_{1s}$   & SCAN/Tight    & 17.73 & 17.73 &  0.25 \\
\ditto     & \ditto                & SCAN/VCC-5Z   & 17.87 & 17.87 &  0.25 \\
\ditto     & \ditto                & TPSS/Tight    & 19.06 & 19.06 &  0.28 \\
\ditto     & \ditto                & TPSS/VCC-5Z   & 19.08 & 19.08 &  0.28 \\
\hline
N $(\#12)$ & $\Delta$-SCF          & SCAN/Tight    & -0.13 &  0.14 &  0.12 \\
\ditto     & \ditto                & SCAN/VCC-5Z   & -0.10 &  0.12 &  0.12 \\
\ditto     & \ditto                & TPSS/Tight    & -0.05 &  0.09 &  0.12 \\
\ditto     & \ditto                & TPSS/VCC-5Z   & -0.02 &  0.09 &  0.12 \\
\ditto     & $-\varepsilon_{1s}$   & SCAN/Tight    & 19.86 & 19.86 &  0.28 \\
\ditto     & \ditto                & SCAN/VCC-5Z   & 19.96 & 19.96 &  0.32 \\
\ditto     & \ditto                & TPSS/Tight    & 21.43 & 21.43 &  0.29 \\
\ditto     & \ditto                & TPSS/VCC-5Z   & 21.45 & 21.45 &  0.30 \\
\hline
O $(\#11)$ & $\Delta$-SCF          & SCAN/Tight    & -0.51 &  0.51 &  0.15 \\
\ditto     & \ditto                & SCAN/VCC-5Z   & -0.41 &  0.41 &  0.15 \\
\ditto     & \ditto                & TPSS/Tight    & -0.43 &  0.43 &  0.15 \\
\ditto     & \ditto                & TPSS/VCC-5Z   & -0.32 &  0.32 &  0.14 \\
\ditto     & $-\varepsilon_{1s}$   & SCAN/Tight    & 22.44 & 22.44 &  0.30 \\
\ditto     & \ditto                & SCAN/VCC-5Z   & 22.45 & 22.45 &  0.30 \\
\ditto     & \ditto                & TPSS/Tight    & 24.05 & 24.05 &  0.30 \\
\ditto     & \ditto                & TPSS/VCC-5Z   & 24.10 & 24.10 &  0.30 \\
\hline
F $(\# 8)$ & $\Delta$-SCF          & SCAN/Tight    & -0.94 &  0.94 &  0.12 \\
\ditto     & \ditto                & SCAN/VCC-5Z   & -0.83 &  0.83 &  0.17 \\
\ditto     & \ditto                & TPSS/Tight    & -0.89 &  0.89 &  0.09 \\
\ditto     & \ditto                & TPSS/VCC-5Z   & -0.75 &  0.75 &  0.13 \\
\ditto     & $-\varepsilon_{1s}$   & SCAN/Tight    & 25.80 & 25.80 &  0.50 \\
\ditto     & \ditto                & SCAN/VCC-5Z   & 25.82 & 25.82 &  0.47 \\
\ditto     & \ditto                & TPSS/Tight    & 27.78 & 27.78 &  0.45 \\
\ditto     & \ditto                & TPSS/VCC-5Z   & 27.82 & 27.82 &  0.47 \\
\hline
\hline
\end{tabular}
\label{tab:scantpss}
\end{table}

For the smallest 32 molecules of \bigqm with 1-3 CONF atoms, excluding oxygen molecule 
which has a triplet ground state, we calculated \cebe 
using the $G_{\Delta H}W_0$ method and  performed
complete basis set extrapolations. Quasiparticle energies were assigned using Mulliken populations, as discussed in \ref{sec_mulliken}. Individual values of \cebe are provided in Table~S4 of SI, which
we use as the reference to assess the accuracy of \cebe  determined using the
mGGA functionals, TPSS and SCAN. The mGGA-DFT methods were selected due to their reduced
computational complexity compared to hybrid-DFT methods, while exhibiting good accuracies for modeling KS quasi-particle energies and
 the bandgaps in solids \cite{kovacs2023origin,ramakrishnan2023bandgaps}.
Furthermore, a previous study has shown that the accuracy of SCAN based \dscf \cite{jana2023slater} can surpass that of $GW$ methods\cite{li2022benchmark}. 
The error metrics for \cebe predicted by
the \dscf approach (Eq.~\ref{eq_1sCEBE}) and the Koopmans' approximation (Eq.~\ref{eq_koopmans}) 
are tabulated in TABLE \ref{tab:scantpss}.

For TPSS and SCAN, the error metrics are similar with their
standard deviations agreeing to $<$ 0.05 eV, suggesting either of the mGGA functionals to be appropriate
for \dscf estimation of \cebe. 
The impact of using frozen orbitals in Koopmans' approximation is apparent from
the severe underestimation ($>$17 eV) of \cebe with the error increasing
systematically with atomic number.
Specifically, for SCAN/Tight-Full, 
the MAE is 17.73 eV for C, which increases to 
19.86, 22.44, and 25.80 eV for N, O, and F, respectively.
The variation between the basis sets when using the same DFT method is negligible, with NAO-VCC-5Z showing slightly lower errors when compared to Tight-Full. 
In some evaluatory calculations, NAO-VCC-5Z resulted in non-convergence of the density for core-ionized cations. Hence, we use Tight-Full for calculating \cebe for \bigqm molecules. 

\begin{figure*}[htbp]
        \centering
        \includegraphics[width=1.0\linewidth]{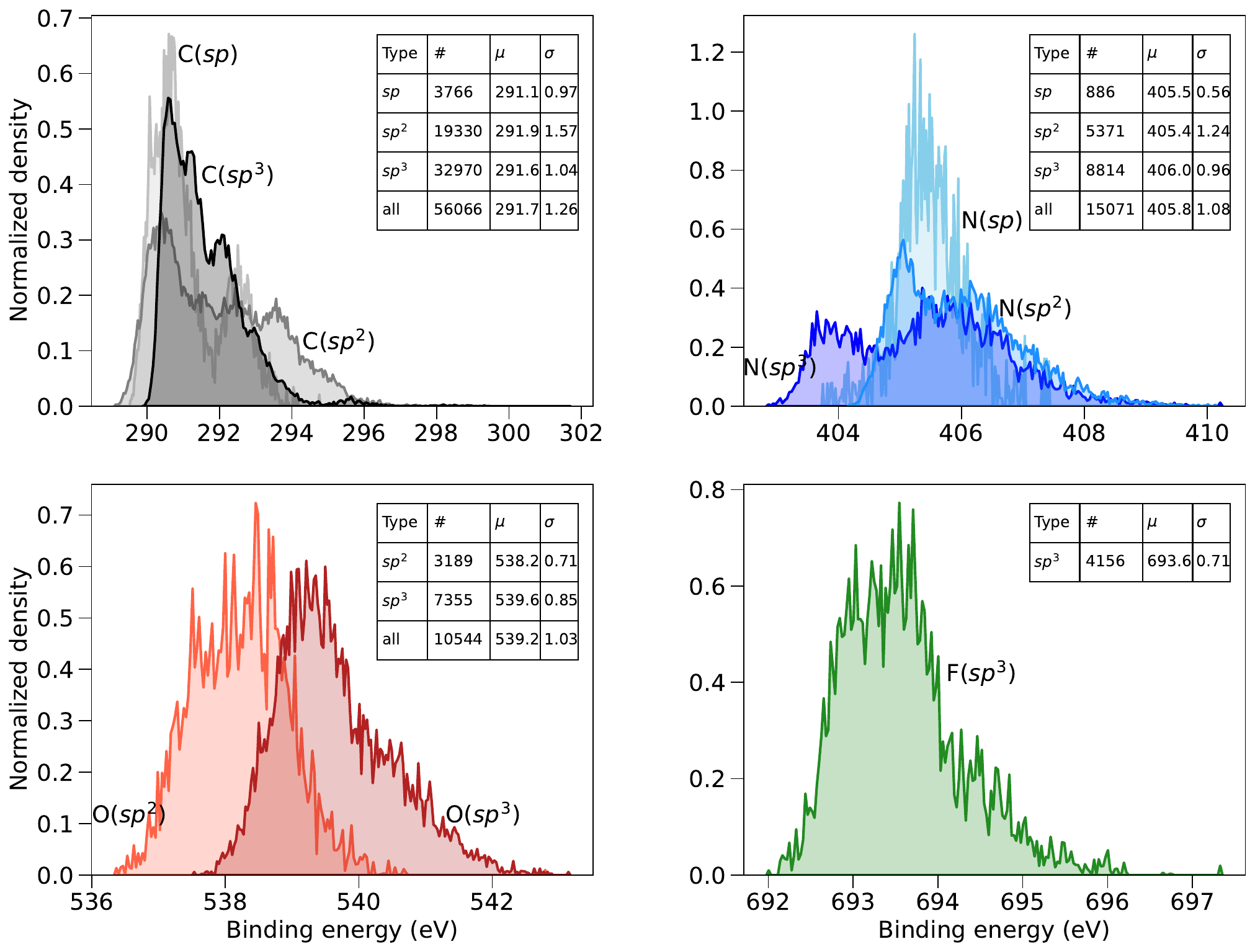}
       \caption{Normalized density plots of CEBEs for distinct hybridizations of CONF 
       atoms in the \bigqm dataset. `$\#$' represents the number of entries, while mean, `$\mu$' and standard deviation, `$\sigma$' are presented in eV. }
       \label{fig:density}
\end{figure*}

\begin{figure*}[htbp]
        \centering
        \includegraphics[width=1.0\linewidth]{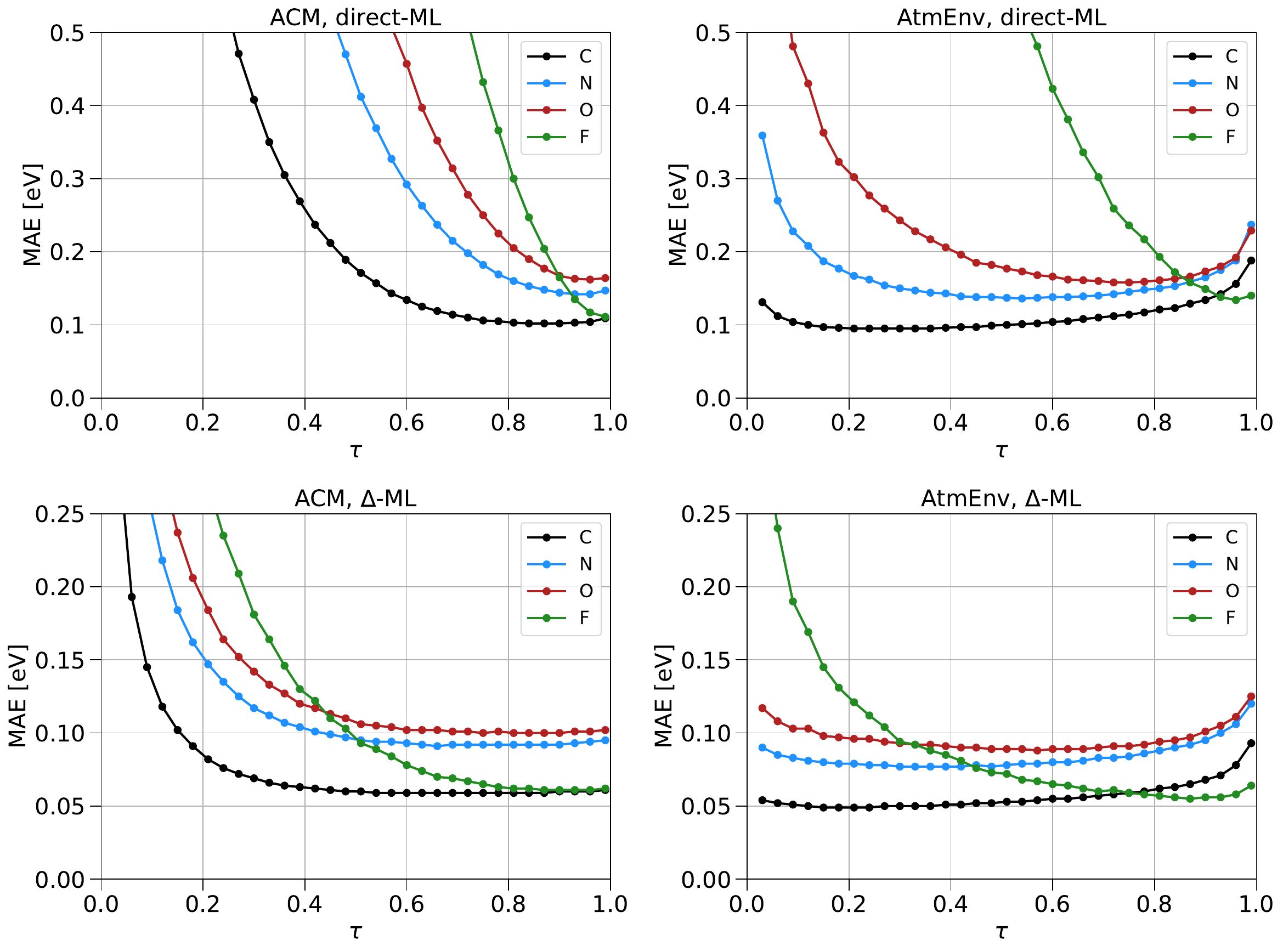}
       \caption{ 
       Variation of MAE (in eV) with $\tau$ for estimating the kernel width ($\sigma$), 
       see Eq.~\ref{eq:sigma_opt_tau} in the main text. Out-of-sample errors for predicting \cebe of C/N/O/F atoms in the bigQM7$\omega$ dataset are shown. Each point on a plot represents the out-of-sample error from an 80/20 train-test split. ACM denotes
       the atomic Coulomb matrix, while AtmEnv denotes the atomic environment determined 
       from a graph neural network framework.
       The scheme direct-ML corresponds to learning on the target \cebe 
       at the SCAN/{\rm Tight-Full} level.
       For $\Delta-$ML, the target quantity is the difference:
       $E_{\rm b}^{\Delta-{\rm SCF}} {\rm (SCAN)}-E_{\rm b}^{\rm Koopmans}(\rm PBE@UFF)$; 
       see the main text for more details. For each element, the {\tt cebeconf} program 
       uses $\tau$ corresponding to smallest MAE.}
       \label{fig:scantau}
\end{figure*}

\begin{figure*}[htbp!]
        \centering
        \includegraphics[width=1.0\linewidth]{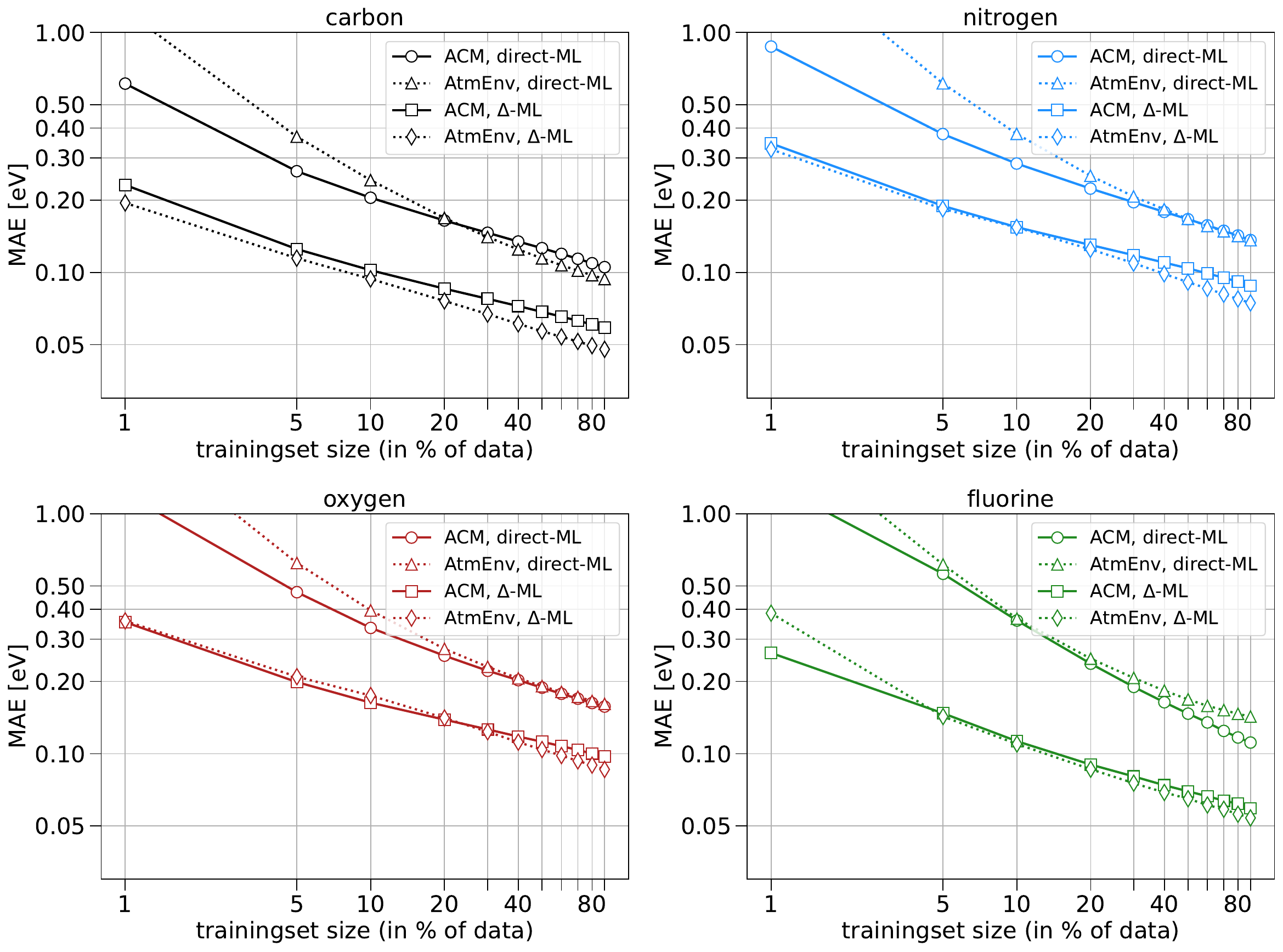}
        \caption{Learning curves showing out-of-sample errors, as a log-log scale plot,
        for KRR-ML for predicting \cebe with increasing training set size. 
        Results are shown separately for C, N, O,
        and F atoms. ACM and AtmEnv correspond to two variants of descriptors
        generated using UFF geometries.
        For direct-ML, the target quantity is 1$s$ CEBE determined using the SCAN/Tight-Full 
        method within the $\Delta$-SCF formalism, while in
        $\Delta$-ML, the model is trained on the difference between the target
        and the baseline estimation of \cebe with in the Koopmans approximation 
        determined using the PBE/cc-pVDZ method using UFF geometries. 
        Each curve was obtained by considering 25 data shuffles, and the mean of MAEs is plotted for various
        training set sizes. The standard deviation of MAEs across shuffles was negligible for large training set sizes. For clarity, we have not shown the envelope.
       }
       \label{fig:learningcurves}
\end{figure*}
\begin{table*}[!hptb]
\caption{ Prediction errors for the direct-ML and $\Delta$-ML
models of \cebe for 208 uracil-type molecules depicted in FIG.~5. Errors are quantified with reference to the target values determined at the SCAN/Tight-Full level of theory within the $\Delta$-SCF formalism.
The input for the ML models are structures determined using the universal force field (UFF). For $\Delta$-ML, 1$s$ energies of core-type molecular orbitals determined at the PBE/cc-pVDZ level using UFF geometries were used as the baseline. 
For a fixed ML model shared through the {\tt cebeconf} module, prediction errors for 208 uracil-type molecules are
reported as the mean absolute error (MAE) and the standard deviation of the errors (SD) in parentheses. Both MAE and SD are in eV. 
Pearson correlation coefficient, denoted by $R$, and the Spearman rank correlation coefficient, denoted by $\rho$, are also included.
Error metrics are provided separately for carbon (C), nitrogen (N), oxygen (O), and fluorine (F) atoms within each set of molecules. The number of atoms of each type present in the set is included in the adjoining parentheses.
The labels P2(1$H$), P4(1$H$), and P4(3$H$) correspond to Pyrimidine-2(1$H$)-one, Pyrimidine-4(1$H$)-one, and Pyrimidine-4(3$H$)-one, respectively. }
\small\addtolength{\tabcolsep}{1.2pt}
\begin{tabular}[t]{ll ll cc cc cc c}
\hline
\multicolumn{2}{l}{Atom} & 
\multicolumn{2}{l}{Molecules} & 
\multicolumn{2}{l}{ML/ACM}  & 
\multicolumn{2}{l}{$\Delta$-ML/ACM}  & 
\multicolumn{2}{l}{ML/AtmEnv} &
\multicolumn{1}{l}{$\Delta$-ML/AtmEnv}  \\
\cline{5-5}
\cline{7-7}
\cline{9-9}
\cline{11-11}
\multicolumn{2}{l}{} &
\multicolumn{2}{l}{} &
\multicolumn{2}{l}{MAE (SD), $R$, $\rho$} &
\multicolumn{2}{l}{MAE (SD), $R$, $\rho$} &
\multicolumn{2}{l}{MAE (SD), $R$, $\rho$} &
\multicolumn{1}{l}{MAE (SD), $R$, $\rho$} \\
\hline
C(72)   &&  uracil    &&  0.53  (0.53),  0.94,  0.95  &&  0.35  (0.45),  0.96,  0.97   &&  0.35  (0.28),  0.98,  0.98 &&  0.20  (0.22),  0.99,  0.99  \\
C(304)  &&  P2(1$H$)  &&  0.63  (0.64),  0.89,  0.90  &&  0.38  (0.48),  0.94,  0.94   &&  0.32  (0.35),  0.97,  0.97 &&  0.25  (0.30),  0.98,  0.98  \\
C(304)  &&  P4(1$H$)  &&  0.55  (0.61),  0.90,  0.90  &&  0.39  (0.48),  0.94,  0.92   &&  0.37  (0.47),  0.94,  0.93 &&  0.22  (0.26),  0.98,  0.98  \\
C(304)  &&  P4(3$H$)  &&  0.49  (0.58),  0.92,  0.92  &&  0.30  (0.33),  0.97,  0.97   &&  0.33  (0.43),  0.96,  0.96 &&  0.12  (0.17),  0.99,  0.99  \\
\hline
N(40)   &&  uracil    &&  0.39  (0.28),  0.81,  0.73  &&  0.22  (0.16),  0.96,  0.96   &&  0.48  (0.38),  0.74,  0.80 &&  0.23  (0.23),  0.92,  0.90  \\
N(176)  &&  P2(1$H$)  &&  0.45  (0.56),  0.90,  0.89  &&  0.19  (0.22),  0.98,  0.98   &&  0.47  (0.45),  0.93,  0.88 &&  0.16  (0.20),  0.99,  0.99  \\
N(176)  &&  P4(1$H$)  &&  0.46  (0.49),  0.95,  0.96  &&  0.20  (0.25),  0.99,  0.99   &&  0.57  (0.58),  0.89,  0.84 &&  0.22  (0.28),  0.99,  0.99  \\
N(176)  &&  P4(3$H$)  &&  0.39  (0.41),  0.90,  0.90  &&  0.18  (0.20),  0.98,  0.98   &&  0.76  (0.59),  0.84,  0.82 &&  0.13  (0.16),  0.99,  0.99  \\
\hline
O(32)   &&  uracil    &&  0.32  (0.33),  0.73,  0.73  &&  0.21  (0.22),  0.95,  0.94   &&  0.26  (0.24),  0.77,  0.74 &&  0.09  (0.10),  0.96,  0.96  \\
O(64)   &&  P2(1$H$)  &&  0.22  (0.30),  0.82,  0.82  &&  0.24  (0.22),  0.94,  0.94   &&  0.51  (0.45),  0.16,  0.12 &&  0.15  (0.18),  0.89,  0.87  \\
O(64)   &&  P4(1$H$)  &&  0.28  (0.25),  0.82,  0.79  &&  0.33  (0.18),  0.96,  0.96   &&  0.49  (0.34),  0.68,  0.66 &&  0.25  (0.13),  0.94,  0.94  \\
O(64)   &&  P4(3$H$)  &&  0.30  (0.38),  0.61,  0.56  &&  0.20  (0.16),  0.96,  0.95   &&  0.29  (0.31),  0.73,  0.71 &&  0.08  (0.09),  0.98,  0.98  \\
\hline
F(8)    &&  uracil    &&  0.40  (0.16),  0.97,  0.98  &&  0.07  (0.07),  0.99,  1.00   &&  0.47  (0.28),  0.95,  0.95 &&  0.15  (0.14),  0.96,  1.00  \\
F(48)   &&  P2(1$H$)  &&  0.35  (0.37),  0.72,  0.69  &&  0.20  (0.09),  0.99,  0.98   &&  0.40  (0.41),  0.79,  0.79 &&  0.14  (0.16),  0.96,  0.97  \\
F(48)   &&  P4(1$H$)  &&  0.25  (0.30),  0.83,  0.82  &&  0.20  (0.12),  0.97,  0.98   &&  0.45  (0.50),  0.62,  0.59 &&  0.18  (0.19),  0.93,  0.93  \\
F(48)   &&  P4(3$H$)  &&  0.32  (0.38),  0.71,  0.68  &&  0.17  (0.12),  0.98,  0.98   &&  0.53  (0.63),  0.10,  0.08 &&  0.22  (0.21),  0.93,  0.93  \\
\hline 
\end{tabular}
\label{tab:ErrorMetrics}
\end{table*}

\begin{figure*}[htbp]
        \centering
        \includegraphics[width=1.0\linewidth]{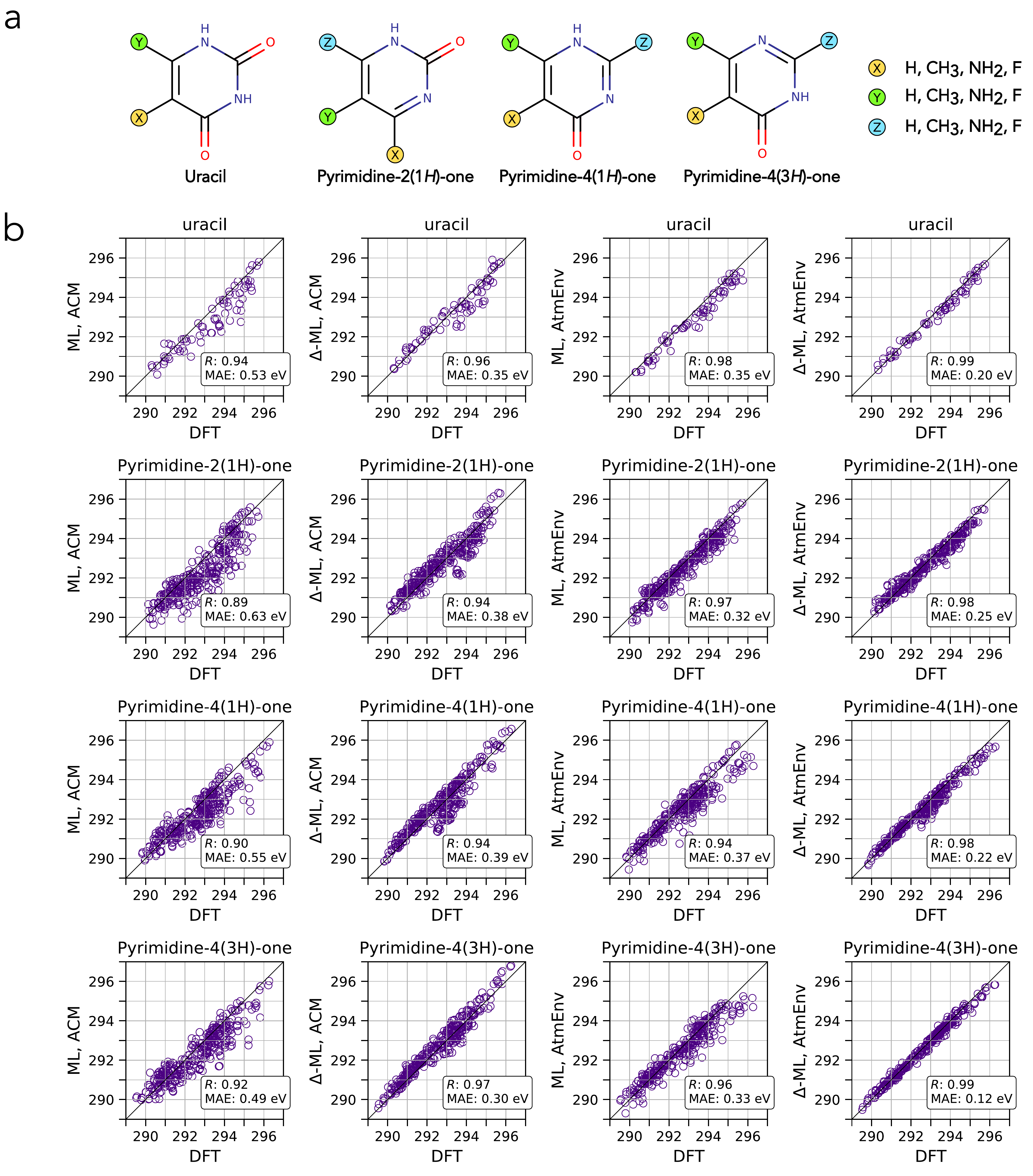}
        \caption{
        a.) Combinatorially varying set of biologically relevant molecules selected for validating the \cebe predicted by the ML models presented in this study. The number of derivatives are Uracil (Pyrimidine-2,4(1$H$,3$H$)-dione): 16, Pyrimidine-2(1$H$)-one: 64, Pyrimidine-4(1$H$)-one: 64, and Pyrimidine-4(3$H$)-one: 64, amounting to 208 unique molecules.
        b.) Scatterplot comparison of ML-predicted and DFT values of \cebe-C in 208 uracil derivatives.
         For all four classes of molecules, ML and $\Delta$-ML predictions are 
         shown along with the 
         Pearson correlation coefficient ($R$) and 
         mean absolute error (MAE).
        }
       \label{fig:scatter}
\end{figure*}

\begin{figure*}[!hptb]
       \centering
       \includegraphics[width=1.0\linewidth]{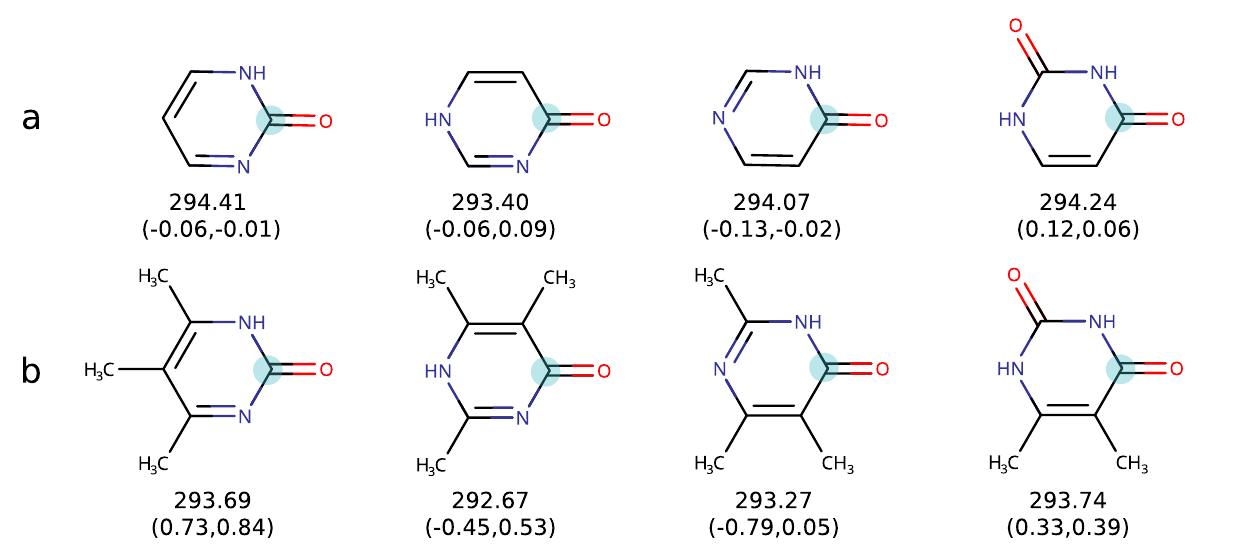}
       \caption{Effect of substitution of methyl groups (-CH$_3$) in the base molecules, 
       shown in row-a, on the \cebe-C of the carbonyl group (blue circle). 
       See the caption of FIG.~\ref{fig:scatter}a for the names of the molecules.
       Corresponding methylated molecules are shown in row-b. 
       SCAN/Tight-Full, $\Delta$-SCF-based \cebe (in eV) of the C atom are given below each molecule. Error in predictions of direct-ML/AtmEnV and $\Delta$-ML/AtmEnv with respect to $\Delta$-SCF values are given in parenthesis.}
       \label{fig_methylatedbase}
\end{figure*}

\subsection{Distribution of 1$s$-CEBEs of CONF atoms}\label{sec_trends}
For an atom in different molecular configurations, \cebe 
is influenced by a combination of the chemical environment provided by the neighboring atoms \cite{greczynski2020x} and screening effects \cite{bagus2013interpretation}. 
Normalized distributions of  \cebe  for different hybridizations of the
CONF atoms in \bigqm are shown in FIG. \ref{fig:density}.
The distribution is heavily biased towards $sp^3$-C with about thirty two thousand
(32k) atoms, constituting about $38\%$ of 85k 
atoms in the dataset. Conversely, $sp$-N is the least represented, 
with only 886 atoms amounting to $0.01\%$ of the dataset. 
The challenge in assigning \cebe lies in the fact that these energies
span narrow ranges with multiple chemical environments 
giving rise to identical values.

In the \bigqm dataset, \cebe of F and C atoms have spreads of
0.71 and 1.26 eV, respectively (see FIG.~\ref{fig:density}). 
Separation into individual hybridizations does not decrease the spread of distribution. In some cases, the standard deviation worsens compared to the unseparated distribution.
The high compositional diversity of the \bigqm dataset renders substructure classification difficult due to the several unique connectivities possible for 
each atom type. 
We analyzed a subset of \bigqm comprising 24 molecules with the
unique substructure: $sp^2$-C  bonded to N via a double bond, while bonded to O and F atoms through single bonds. See Figure~S3~a) of the SI for the 
 corresponding molecular structures. 
All molecules in the subset are 5-membered heterocyclic rings containing 
C, N, and O atoms that define the substructure. 
For the 25 F atoms in this molecular set, the distribution of \cebe 
span a range of 1.9 eV with a 
standard deviation of 0.5 eV, 
only 0.2 eV lower than that of all 4156 F atoms in the \bigqm dataset. 
A comparison of the distributions is shown in  Figure~S3~b) of the SI. This 
indicates that partitioning $E_{\rm b}^{1s}$ of CONF atoms in \bigqm 
based on substructure similarity does not decrease the spread of the values,
and the scope for generating separate ML models based on substructure
classification is limited.

\subsection{Optimization of ML models} 
\label{sec_eval}
We probed the impact of the length of AtmEnv descriptor 
vector on the accuracy of
KRR-ML models of \cebe of C atoms.
Results, detailed in Table~S3 of the SI, show that the out-of-sample prediction accuracy improves marignally going from a length of 128 to 1024 for direct-ML predictions.
Given an error of approximately 0.1 eV at length 128, the errors for lengths 32 and 64 are likely to exceed 0.1 eV. 
For $\Delta$-ML, the accuracy remains at 0.05 eV for length 128, slightly decreasing by only 7 meV at length 1024. 
Our goal of achieving an MAE below 0.1 eV for data-driven predictions of \cebe while keeping the computational complexity minimal led us to limit the length of
AtmEnv to 128.

In order to visualize AtmEnv obtained from SchNet, we considered
six classes of molecules not necessarily contained in \bigqm: 
(a) CH$_3$-R, (b) CH$_3$-CH$_2$-R, (c) H$_2$C=C(H)-R, (d) HC$\equiv$C-R, 
(e) substituted benzene (C$_6$H$_5$-R), and (f) substituted 
cyclohexane (C$_{12}$H$_{11}$-R), 
where R = H, CH$_3$, NH$_2$, OH, and F. 
We used a trained SchNet model to generate AtmEnv 
for the `C' atom marked and plotted the heatmaps shown 
in Figure~S4 of the SI. 
The variations in the elements of AtmEnv, associated with groups bonded to C, demonstrate how SchNet embeddings capture atomic environments. Although embeddings are not physically interpretable, they encapsulate atomic environments \cite{hu2023aisnet,dybowski2020interpretable}. For a fixed length of the AtmEnv vector, identical values for a given element of the vectors of different molecules indicate the presence of similar chemical environments in these molecules.

To determine the optimal kernel width, $\sigma_{\rm opt}$, for KRR-ML,
we varied the parameter $\tau$ between 0 and 1 and applied Eq.~\ref{eq:sigma_opt_tau}.
For each atom type, we used randomly selected 80\% of the data to determine the hold out error on the remaining 20\%. Further, we used same indices across different ML/descriptor combinations: direct-ML/ACM, direct-ML/AtmEnv, $\Delta$-ML/ACM, and $\Delta$-ML/AtmEnv. 
 FIG.~\ref{fig:scantau} displays the variation in MAE with  $\tau$ for all atom types. $\Delta$-ML requires a 
 smaller $\tau_{\rm opt}$ (hence larger $\sigma_{\rm opt}$ as per Eq.~\ref{eq:sigma_opt_tau}) than direct-ML for all atoms. This indicates the optimal kernel function for direct-ML to be broader, sampling over more training examples. 
 The sensitivity of MAE to $\tau$ increases from C to F atoms. 
  While $\sigma_{\rm opt}$ determined through scanning $\tau$ for \cebe-C are comparable to those
 determined using the heuristics proposed in a previous study\cite{ramakrishnan2015many} with $\tau=1/2$ (see Eq.~\ref{eq:sigma_opt_max}), precise tuning becomes crucial for modeling \cebe of F. The optimal values of $\sigma_{\rm opt}$ determined using 
$\tau$ corresponding to the minimum MAE in  FIG.~\ref{fig:scantau} and  Eq.~\ref{eq:sigma_opt_tau}
are hardcoded in the trained ML models provided in the module {\tt cebeconf}\cite{cebeconf}. The target accuracy (MAE $<$ 0.1 eV) for obtaining distinct CEBEs, for resolving XPS is achieved adequately by our $\Delta$-ML models. Figure S5 in the SI illustrates XPS predictions for a test molecule using these pre-trained models via \texttt{cebeconf}.

\subsection{Performance of ML models for out-of-sample predictions} \label{sec_learning}

FIG.~\ref{fig:learningcurves} presents learning curves 
depicting the MAEs for out-of-sample prediction
as a function of the training set size for 
various KRR-ML models.
The monotonous drop in MAE with the training set implies 
that the models capture structure-property correlation that improve with increasing example data.
Overall, across CONF elements, ACM and AtmEnv show similar learning trends for direct-ML predictions. 
While using 90\% data of \cebe-C for training in direct and $\Delta$-ML, AtmEnv performs 
slightly better than ACM. Only in the case of
direct-ML for \cebe-F, ACM converges to better accuracy than AtmEnv. Direct-ML predictions based on ACM and AtmEnv offer similar accuracies for N and O. 
In $\Delta$-ML, AtmEnv delivers consistently lower MAE compared to ACM for modeling 1$s$-CEBEs of all elements.

As discussed in  \ref{sec_trends}, \cebe of C is the most represented 
consisting of $\approx65\%$ of the \bigqm-CEBECONF dataset, achieving $\Delta$-ML MAE of 0.05 eV with AtmEnv descriptor. 
Additionally, the MAE for \cebe-F when using 
the $\ge80\%$ of the data for training drops below 0.07 eV. 
However, a look at the distribution and composition of the F set offers insight into this high accuracy. Apart from the possibility of a single hybridization, all F atoms are connected to $sp^3$ C atoms, except the outliers HF and F$_2$, making the set compositionally more uniform compared to N and O. Hence, the distribution of \cebe-F also has a lower standard deviation compared to C/O/N, see \ref{fig:density}. 
While N and O groups have a higher compositional diversity compared to F, their training sizes are not large compared to C, resulting in higher MAEs. 
Overall, for data-driven predictions of \cebe of C/O/N/F atoms
to an accuracy of $\le0.1$ eV, $\Delta-$ML approach is necessary; the target accuracy is reached with 10\%, 40\%, 60\%, and 20\% of the data for C, N, O, and F
atoms, respectively. Decreasing the error further may require better baselines in $\Delta$-ML or better geometries, both may incur additional
expenses for out-of-sample predictions.

\subsection{Transferability Test for ML models}\label{sec_validation}
Since \cebe is a property of an atom in a molecule, it is of interest to 
explore the transferability of the ML models trained on \bigqm to new class of molecules.
Hence, we have selected a set of aromatic heterocyclic systems as a validation set. 
    Its composition is motivated by the potential application of the ML-models of \cebe 
    to identify the 
    composition of small biomolecules such as derivatives of nucleic acid bases, heavily substituted by O and N. 
    Additionally, the use of substituted pyrimidines for synthesis is common \cite{parker2009enzymology}. 
    Particularly, up to two carbonyl groups have been included in pyrimidines 
    generating derivatives of  
    uracil, pyrimidine-2(1H)-one,  pyrimidine-4(1H)-one and pyrimidine-4(3H)-one, see FIG. \ref{fig:scatter}a. 
    C atoms not attached to carbonyls are combinatorially substituted with either H, CH$_3$, NH$_2$ or F.

    Each of the three pyrimidinones has a carbonyl group and three sites for substitution, leading to 64 derivatives in each class, summing to 192 molecules. Additionally, 16 substituted uracils are generated, since there are 2 carbonyls attached to a pyrimidine, availing the remaining 2 C atoms for substitution. These molecules contain 7-10 CONF atoms. The number of molecules containing 7,8,9 and 10, CONF atoms, 
    are 1, 9, 27 and 27, respectively in each of the pyrimidinone sets. The uracil set has 1,6 and 9, molecules with 8,9, and 10 CONF atoms, respectively.
    Together, these 208 molecules constitute our validation set. Among them, 
    only one (unsubstituted pyrimidine-2(1H)-one) is contained in \bigqm. Along with this molecule,
    pyrimidine-4(1H)-one and pyrimidine-4(3H)-one are the only molecules 
    in the validation set with 7 heavy atoms, while the remaining 206 molecules are composed of more than 7 CONF atoms. 
    Baseline energies for $\Delta$-ML are obtained at the same level of theory as in the training set, 
    with aZORA corrected KS-eigenvalues obtained with the PBE/cc-pVDZ method using UFF-level molecular structures. 
    One can also use the non-relativistic variant of these energies and correct them by adding 
    0.29/0.55/0.97/1.57 eV for C/O/N/F atoms, respectively as stated in Figure~S1.

ML-predicted CEBEs alongside DFT values are represented in scatter plots for all the ML models featured in FIG. \ref{fig:scatter}b. 
Corresponding atom-specific errors 
for each class of molecules are tabulated in TABLE~\ref{tab:ErrorMetrics}. 
Alike out-of-sample predictions in \bigqm, discussed in \ref{sec_learning}, the performance of direct-ML is inferior to that of $\Delta$-ML. 
In general, a correlation coefficient above 0.7 is considered to indicate a strong linear correlation in different disciplines\cite{ratner2009correlation, akoglu2018user}. In this study, all $\Delta$-ML models achieve a high Pearson's correlation coefficient, $R \approx 0.9$, for all classes of validation molecules, as shown in FIG.~\ref{fig:scatter} 
and TABLE~\ref{tab:ErrorMetrics}. Similarly, all direct-ML models exhibit correlation coefficients of $\gtrsim$ 0.7, with the exception of two cases where the correlation is weak.
ACM models perform slightly worse, with direct-ML models based on 
ACM performing the poorest of all models, $\Delta$-ML/ACM has performance similar to direct-ML/AtmEnv predictions.
$\Delta$-ML with AtmEnv shows the least errors, with $R\ge$ 0.98. 
In terms of element-specific accuracy, ACM has better prediction for N and F systems.  
Compared to explorations with the \bigqm dataset, the accuracy of ML models has somewhat deteriorated while applying to large biomolecules. 
This is due to the fact that the molecules in the validation set are larger than those used for training the models. 
However, the qualitative trends between \cebe and the structural variations is captured by our ML models, 
indicated by the high correlation coefficients.

In FIG.~\ref{fig_methylatedbase}, we compare ML-predicted \cebe with that of DFT values for 
the four unsubstituted bases shown in FIG.~\ref{fig:scatter}a
and their CH$_3$ substituted counterpart shown in FIG.~\ref{fig:scatter}b. In particular, we explore the CEBEs of the
C atom in a carbonyl group common to all four base molecules. 
Both direct- and $\Delta$-ML predictions show excellent accuracy 
for unsubstituted pyrimidine-2(1H)-one,
which is contained in the \bigqm dataset. 
For all four unsubstituted molecules, the $\Delta$-ML
predictions show an error of $<0.1$ eV.
Upon methylation, the CEBE at the carbonyl C decreases according to DFT predictions, 
while there is a drastic depreciation in the prediction accuracy of ML models. 
All predictions for the substituted molecules have errors above 0.3 eV except a single case of $\Delta$-ML prediction, where the the prediction is within an error of 0.05 eV. The same C
with direct ML doesn't show a similar drop in error, 
even though the training data is same for both direct-ML and $\Delta$-ML. 
Overall, this
test of transferability of $\Delta$-ML/AtmEnv based on an
inexpensive baseline suggests that CEBEs can be assigned to 
class of molecules with systematic variation of structure 
and composition with high degree of linear correlation.
However, the models do not offer quantitative accuracy for the 
prediction of \cebe when the application is limited to a few molecules, 
as the models based on local descriptors do not capture long-range 
stereo-electronic effects resulting in accumulation of systematic errors.

\section{Conclusion}\label{sec_conclusion}
In this study, we showcased the applicability of ML models 
for prediction of XPS. Using CEBEs determined using \dscf calculations, we 
generated the database, \bigqm-CEBECONF, consisting of 12679 molecules from the \bigqm dataset amounting to 85837 1$s$-CEBE of CONF atoms. For ML modeling, 
we explored the use of two descriptors: ACM, a local version of Coulomb matrix, and AtmEnv, a representation obtained from atom-specific embeddings of a graph neural network framework. 
KRR models were optimized by tuning the kernel widths determined specifically for each atomic model. 
The ML models reported in this study bypass the use of many body calculations or charge constraining algorithms, using only structural description of the molecule for predicting CEBEs. 
These models have been made accessible for general use through the python module {\tt cebeconf}\cite{cebeconf}, where UFF level structures can be given as input to get CEBEs. 
The accuracy is leveraged further with the use of $\Delta$-ML using inexpensive baseline KS eigenvalues from single-point calculation on molecular geometry determined with force-fields.

For out-of-sample predictions within the \bigqm dataset, 
$\Delta$-ML in combination with the AtmEnv descriptor
offers an accuracy of less than 0.1 eV, which is an order of 
magnitude smaller than the spread of 1$s$-CEBE 
in the \bigqm dataset. To understand the transferability of the models, 
we applied them to 
208 biomolecules with homogeneously varying structure and compositions. 
Since these molecules are larger than the
\bigqm molecules used for training, their accuracy dropped 
compared to out-of-sample predictions within the
\bigqm dataset. 
However, they yield an overall correlation coefficient of over 0.9, underscoring the crucial role of correlation in indicating that the prediction errors are largely systematic. This high correlation coefficient suggests that the model may be suitable for large-scale predictions, as it can be effectively calibrated using a few CEBEs determined at the target level.

This study extends the applicability of currently available ML models for XPS prediction to organic molecules. Such models of quasi-atomic properties can be extended to atoms beyond CONF, such as alkali and alkaline elements\cite{kahk2023combining}, transition metal atoms\cite{jorstad2022delta}, and others\cite{hirao2021koopmans}. Currently, large datasets needed for ML modeling for other elements are lacking. Experimental references\cite{jolly1984core} can serve as a starting point for selecting computational chemistry methods that can be employed for data generation.

\section{Supplementary material}
See the supplementary material for the following: 
Figure S1 shows linear fits to obtain aZORA corrections for non-relativistic baseline energies. 
Figure S2 presents the choice of atomic indices in ACM representation on a schematic molecule. 
Figure S3 displays structures and normalized density plots for molecules with an identical substructure. 
Figure S4 shows the heatmaps of AtmEnv of `C' atom in six classes of molecules. 
Figure S5 shows the use of \texttt{cebeconf}.
Table S1 shows the splitting of CEBEs in benzene.
Table S2 gives the composition of the \bigqm-CEBECONF dataset. 
Table S3 shows the variation of MAE for direct-ML and $\Delta$-ML with the length of AtmEnv. 
Table S4 provides the G$_{\Delta H}$W$_0$ energies at cc-pVnZ and extrapolated values at CBS limit. 
Table S5 Table S5 provides statistics of molecules exhibiting convergence failure.
Table S6 gives the list of SMILES of molecules exhibiting convergence failure.
The trained ML models are provided along with example validations at 
\href{https://github.com/moldis-group/cebeconf}{\tt https://github.com/moldis-group/cebeconf}\cite{cebeconf}. 

\section{Data Availability}
The data that support the findings of this study are
within the article and its supplementary material.

\section{Acknowledgments}
We thank Miguel Caro for providing the geometries of a
subset of QM9 used in \RRef{golze2022accurate}. 
We thank Stijn De Baerdemacker for sharing the preprint of \RRef{el2024global}. 
ST thanks Dorothea Golze for commenting on questions regarding $GW$ calculations. 
We thank Atreyee Majumder for assistance in calculations.
We acknowledge the support of the Department of Atomic Energy, Government of India, under Project Identification No.~RTI~4007. 
All calculations have been performed using the Helios computer cluster, which is an integral part of the MolDis Big Data facility, TIFR Hyderabad \href{http://moldis.tifrh.res.in}{(http://moldis.tifrh.res.in)}.

\section{Author Declarations}
\subsection{Author contributions}
\noindent 
{\bf ST}: 
Conceptualization (equal); 
Analysis (equal); 
Data collection (main); 
Writing (equal).
{\bf SD}: 
Analysis (equal); 
Writing (equal).
{\bf SJ}: 
Analysis (equal); 
Writing (equal).
{\bf RR}: Conceptualization (equal); 
Analysis (equal); 
Funding acquisition; 
Project administration and supervision; 
Resources; 
Writing (equal).
\subsection{Conflicts of Interest}
The authors have no conflicts of interest to disclose.

\section*{References}
\bibliography{ref} 
\end{document}


\begin{figure}
    \centering
    \includegraphics[width=\textwidth]{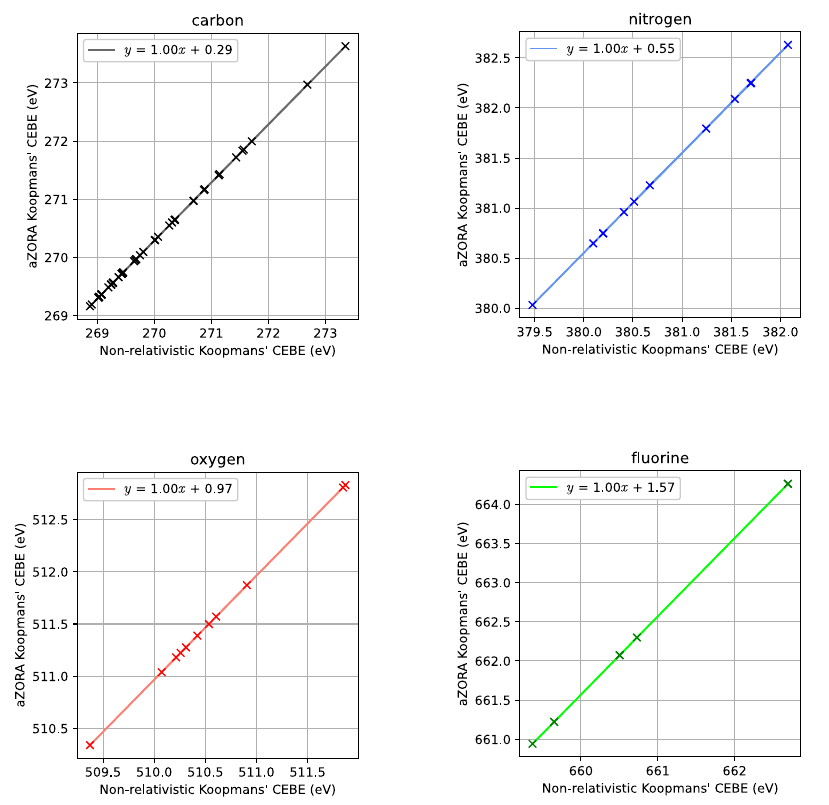}
    \caption{
    Systematic effect of aZORA correction on 1$s$-CEBEs (\cebe) of CONF atoms 
    shown by straight-line fitting
    of non-relativistic values to aZORA values. The effect is shown for the baseline for
    $\Delta$-ML, 
    Koopmans' predicted
    \cebe for the first 32 molecules (excluding O$_2$) from bigQM7$\omega$ 
    at the PBE/cc-pVDZ level using geometries at the UFF level.
    As the slopes in all plots are nearly unity, as expected for atom-level corrections, 
    aZORA values can be obtained using non-relativistic values by adding the intercept
    0.29/0.55/0.97/1.57 eV
    for C/O/N/F atoms, respectively. 
    }
    \label{fig:aZORA_correction}
\end{figure}

\begin{figure}[htbp]
        \centering
    \includegraphics[width=0.75\linewidth]{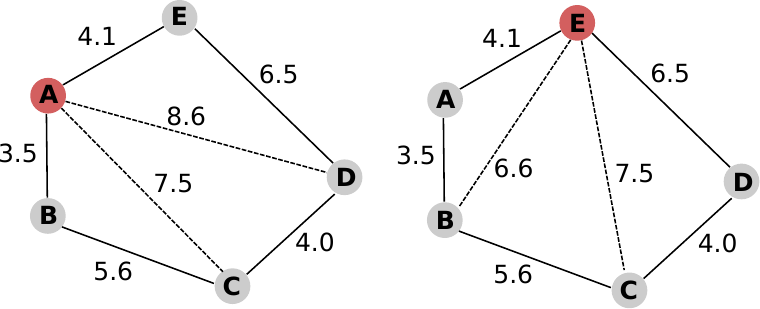}
        \caption{
        Sorting of atomic indices in the ACM (atomic Coulomb matrix) 
        representation for a schematic
        molecule. For query atoms shown in red circles, 
        A (left) and E (right), neighboring atoms with 
        increasing interatomic distances are
        [B, E, C, D] and [A, D, B, C], respectively.
        }
        \label{fig:acm}
\end{figure}

\begin{figure*}[htbp]
    \centering
    \includegraphics[width=\textwidth]{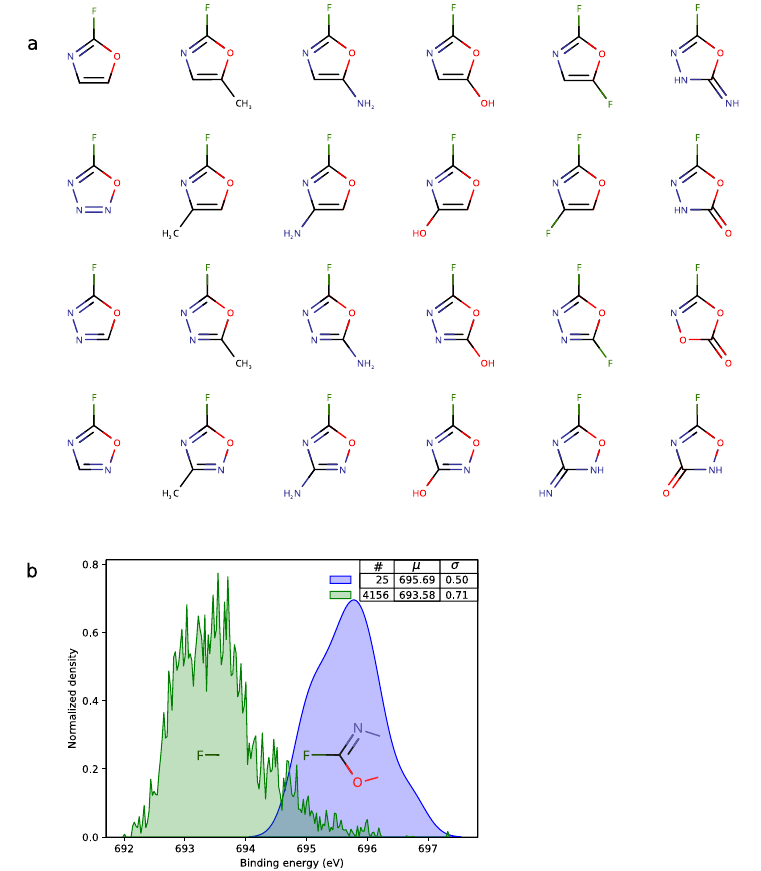}
    \caption{
    a) Five membered molecules with F atoms in the identical
    atomic environment, {F-C(O-)(=N-)}.   
    b) Distributions of \cebe are shown for 
    all F atoms in bigQM7$\omega$ (in green, bin width 0.05) and
    F atoms in the subset of bigQM7$\omega$ with the common substructure 
    shown in panel-a (in blue, bin width 1.0).
    In the legend, $\#$, $\mu$, and $\sigma$ indicate the count, mean (in eV), and standard deviation (in eV), respectively. The area under each density curve integrates to \#.
    }
    \label{fig:N2O1}
\end{figure*}

\begin{figure*}[htbp]
       \centering
       \includegraphics[width=1.0\linewidth]{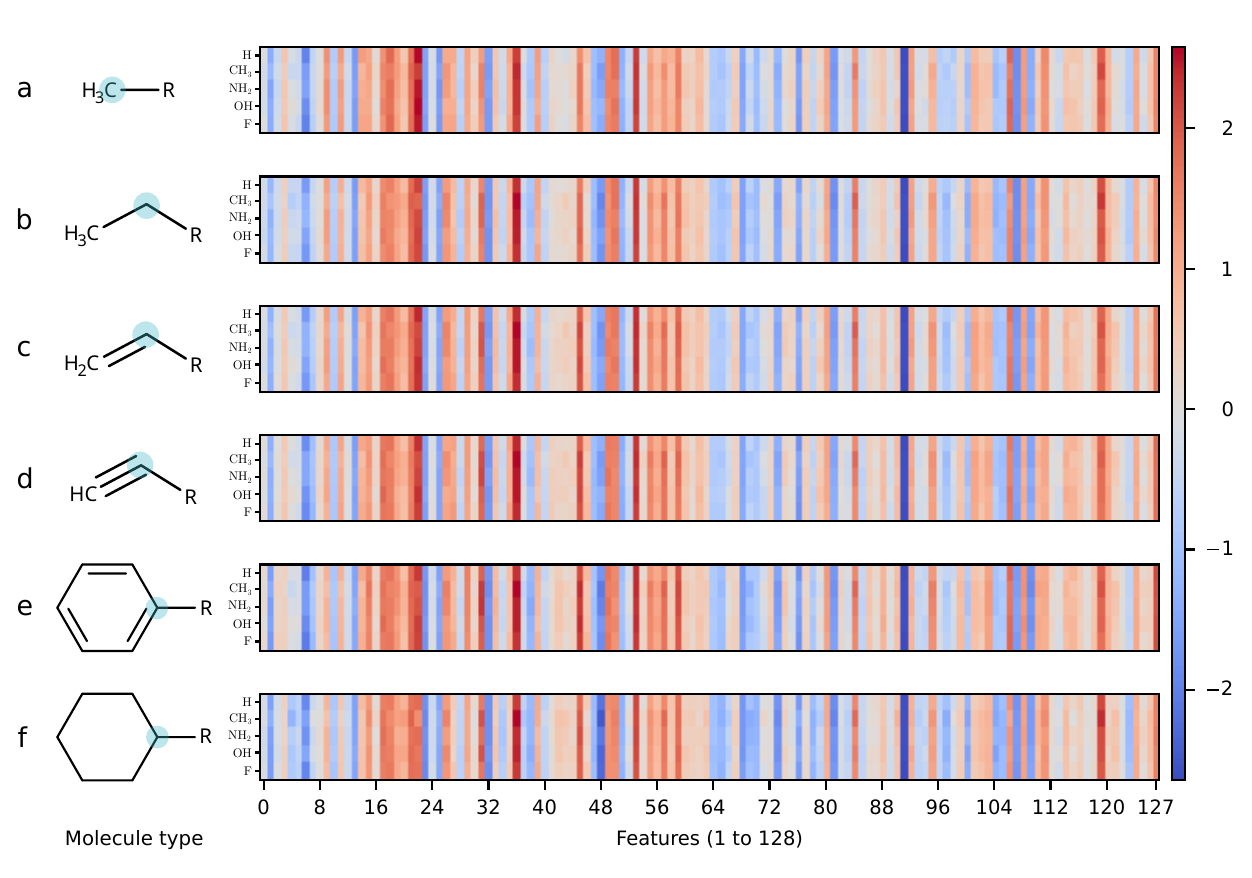}
       \caption{
       For six representative classes of molecules 
       AtmEnv vectors of the highlighted `C' atom are
       color coded:
       a) methane (CH$_3$-R),
       b) ethane (CH$_3$CH$_2$-R),
       c) ethylene (CH$_2$CH-R),
       d) acetylene (CHC-R),
       e) benzene (C$_6$H$_5$-R), and
       f) cyclohexane (C$_6$H$_{11}$-R). 
       For each class of molecule, five derivatives
       are considered with R=-H, -CH$_3$, -NH$_2$, -OH, and -F, and
       AtmEnv vectors are plotted as five rows.
       Identical values for a given element of the vectors of different molecules indicate the presence of similar chemical environment.
       }
       \label{fig:data}
\end{figure*}

\begin{figure}[htbp]
    \caption{
    Screenshots demonstrating the use of \texttt{cebeconf} within a Jupyter notebook. This notebook is provided in the project's repository in the directory \href{https://github.com/moldis-group/cebeconf/tree/main/example\_query}{\tt example\_query}.
    a) Install \texttt{cebeconf} and load all the pre-trained models using the function \texttt{calc$\_$be}; 
    b) and c) show direct-ML predictions (keyword `direct') for ACM and AtmEnv representations, using keywords `ACM' and `AtmEnv' respectively, for a geometry `test.xyz'; 
    d) and e) show $\Delta$-ML predictions (keyword `delta') for ACM and AtmEnv, respectively,
    for which baseline energies at PBE/cc-pVDZ level must be provided as a list in the input in the same order as the CONF atoms in the xyz file. In a), b), c), and e), the code produces an output spectrum by summing Gaussian functions at the predicted binding energies. 
    $\Delta$-ML predictions can also be made without providing baseline values as input if required, as shown in f), and the baseline values can be added separately, as shown in g) to estimate CEBEs at the target level. For brevity, we have omitted the
    module's header from the output.
    }
    \centering
    \includegraphics[width=1.0\linewidth]{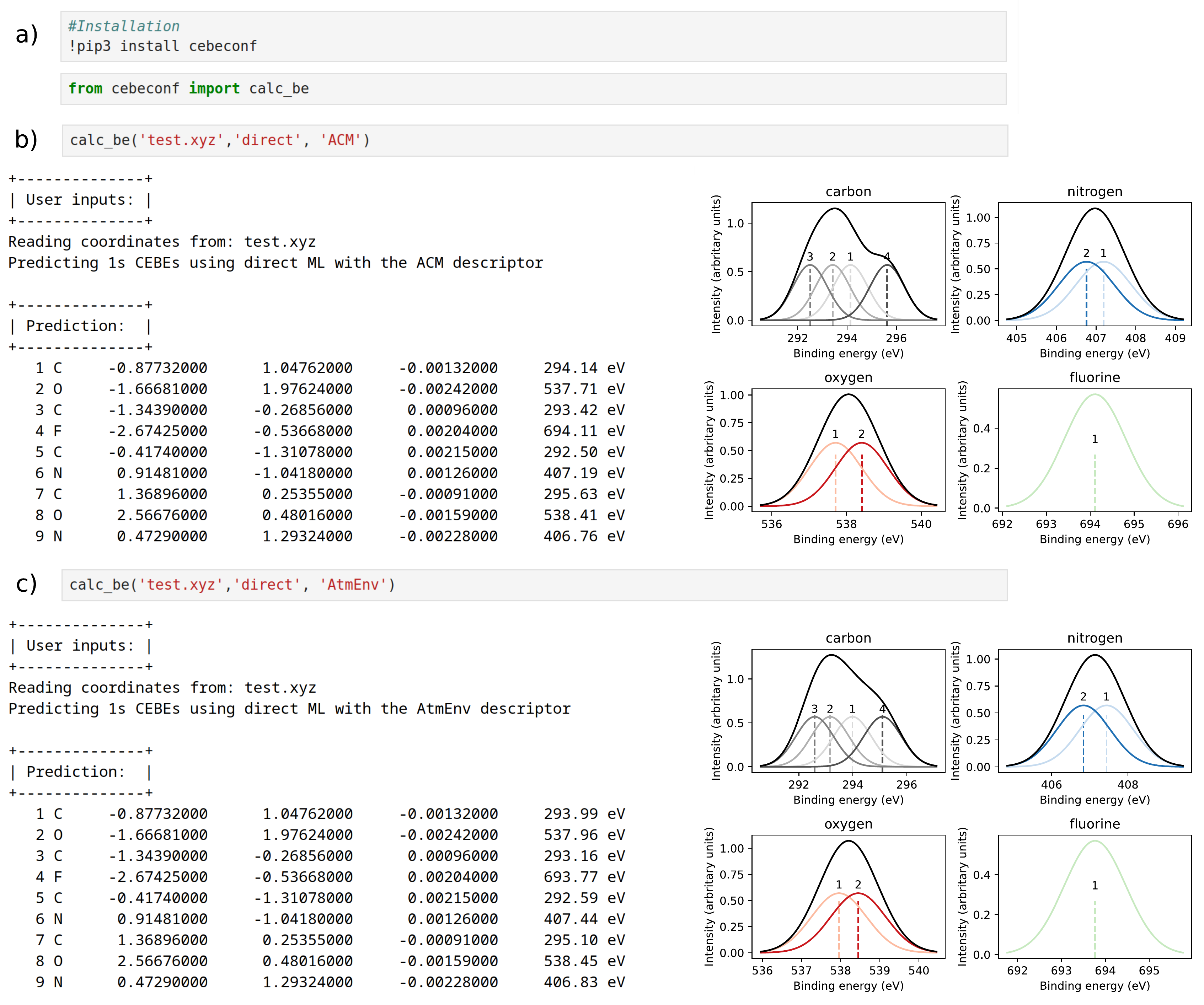}
    \label{fig:enter-label}
\end{figure}
\begin{figure}[htbp]\ContinuedFloat
    \centering
    \includegraphics[width=1.0\linewidth]{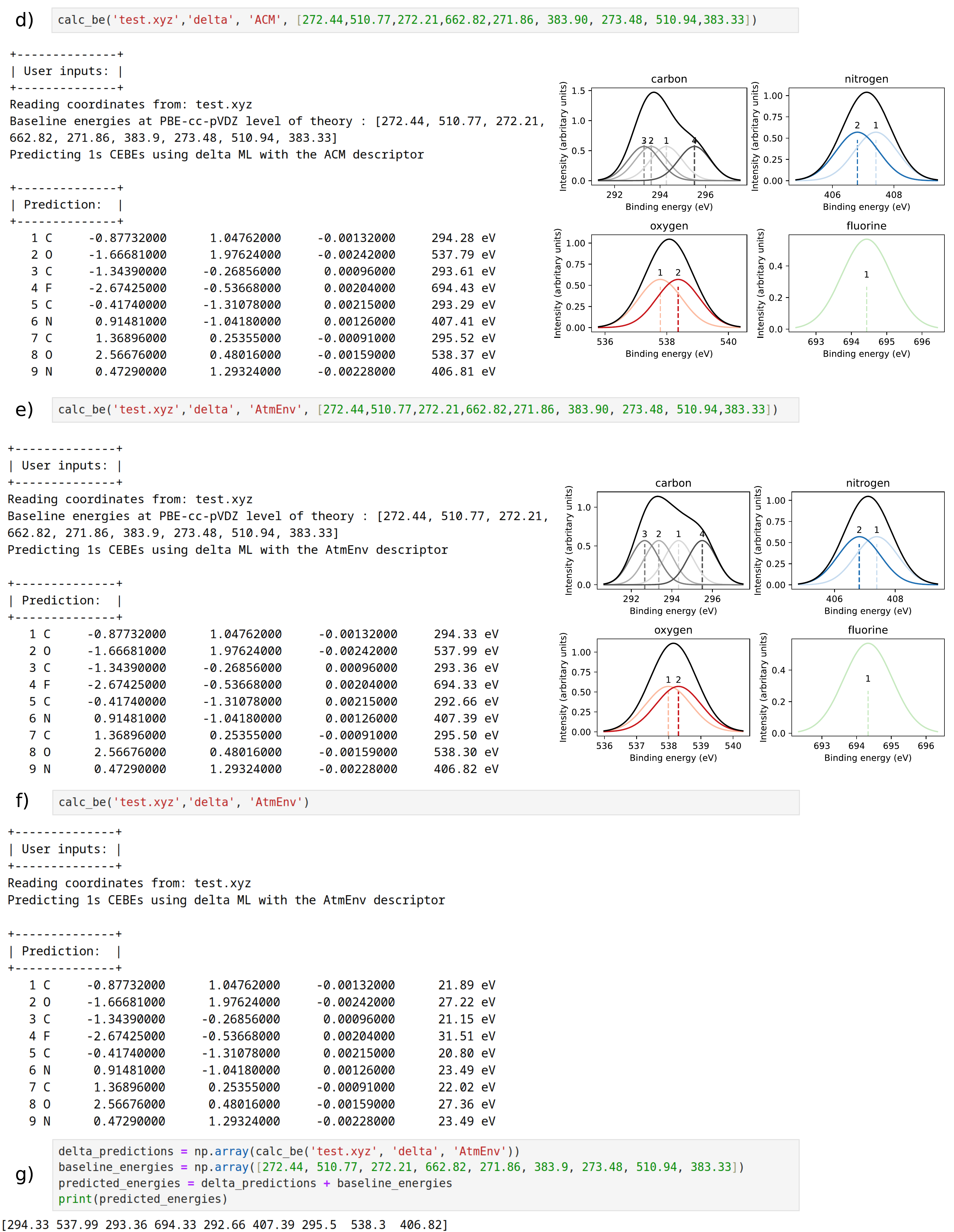}
    \label{fig:enter-label}
\end{figure}

\clearpage
\begin{table}[htbp]
    \centering
    \begin{tabular}{l c c c c }
    \cline{2-5}
      & Koopmans & G$_{\Delta H}$W$_0$ & $\Delta$-SCF & Expt. \cite{myrseth2002vibrational} \\
     \cline{2-5}
            & 0.023 & 0.105 & 0.000 & 0.064 \\
            & 0.017 &  0.034 & 0.000 & 0.048 \\
            & 0.017 &  0.034 & 0.000 & 0.048 \\
            &  0.006 &  0.023 & 0.000 & 0.016 \\
            &  0.006 &  0.021 & 0.000 & 0.016 \\
            &  0.000 &   0.000 & 0.000 & 0.000 \\
     \cline{2-5}
     base & 269.29 & 290.09 & 290.29  & 290.26 \\
     \cline{2-5}
    \end{tabular}
    \caption{Splitting of six \cebe values in benzene (in eV) calculated using Koopmans, G$_{\Delta H}$W$_0$, and $\Delta$-SCF approximations. For reference, the final column includes the 
    experimental data\cite{myrseth2002vibrational}. 
    The deviations are calculated from 
    the minimum values (base) collected in the bottom row. 
    $\Delta$-SCF calculation is performed with SCAN/Tight-Full, whereas Koopmans and GW calculations are performed at the PBE/cc-pV5Z DFT level.}
    \label{tab:benzene_assignment}
\end{table}

\begin{table}[H]
\centering
\caption{Number of molecules 
with C, N, O, F atoms in the bigQM7$\omega$ dataset 
along with the number of entries (\#) of \cebe. 
The number of molecules converged in $\Delta$-SCF 
calculations is also given.}
\begin{tabular}[t]{l c c c  c}
\hline
Element & Molecules & Converged Molecules & \#\cebe \\
\hline
C & 12875 & 12674 & 56066 \\
N &  8486 &  8368 & 15071 \\
O &  7774 &  7628 & 10544 \\
F &  3256 &  3230 & 4156  \\
\hline
all & 12880 & 12679 & 85837 \\
\hline
\end{tabular}
\label{tab:mol_bigqm}
\end{table}%

\begin{table}[htbp]
\centering
\caption{Variation of out-of-sample MAE (in eV) 
for direct-ML and $\Delta$-ML
with the length of AtmEnv feature vector. 
}
\begin{tabular}[t]{ll ll cc cc cc c}
\hline
\multicolumn{2}{l}{} & 
\multicolumn{6}{c}{MAE (eV)}  \\
\hline
Length        &&  128    && 256    && 512 && 1024\\
direct-ML        &&  0.097  &&  0.087 && 0.080 && 0.076\\
$\Delta$-ML   &&  0.050  && 0.046 && 0.044 && 0.043\\
\hline 
\end{tabular}
\label{tab:dir_del}
\end{table}

\clearpage

\begin{longtable}{llll llll}
\caption{$G_{\Delta H}W_0$ estimated \cebe using cc-pVnZ (n=T,Q,5) basis sets,
and values extrapolated at the complete basis set (CBS) limit.
Accuracy of a straight line fitted to the cc-pVnZ (n=T,Q,5) energies compared
to CBS are quantified through mean absolute error (MAE) and coefficient of
determination, $R^2$. 
In the cases where the trend in binding energies is non-monotonic, we have taken the BE obtained using cc-pV5Z instead of CBS energy, denoted by a $\dagger$. 
The first and last atom of three-membered cyclic molecules (last 9 entries)
are linked to each other, this is denoted by -. 
For chemically equivalent atoms in a molecule,  
the KS state denoted with *, corresponding to maximum Mulliken population, is selected as the reference in TABLE I. See, IIB of the article for more details.}\\
\hline 
Molecule & Atom & cc-pVTZ & cc-pVQZ & cc-pV5Z & CBS & MAE(eV) & $R^2$ \\
\hline 
CH$_4$ & C                        & 290.201 & 290.479 & 290.610 & 290.761 & 0.00 & 1.00 \\
NH$_3$ & N                        & 404.703 & 405.367 & 405.398 & 405.766 & 0.07 & 0.94 \\
H$_2$O & O                        & 539.528 & 539.630 & 539.506 & 539.505$\dagger$      \\
HF     & F                        & 694.123 & 694.014 & 693.922 & 693.852 & 0.01 & 0.97 \\
H$_3$C-CH$_3$ & C                 & 290.093 & 290.381 & 290.510 & 290.671* & 0.00 & 1.00 \\
&                                 & 290.079 & 290.391 & 290.506 & 290.680 & 0.01 & 1.00 \\
H$_2$N-CH$_3$ & C                 & 290.880 & 291.205 & 291.164 & 291.164$\dagger$      \\
H$_2$N-CH$_3$ & N                 & 404.765 & 404.941 & 405.060 & 405.136 & 0.02 & 0.97 \\
HO-CH$_3$ & C                     & 291.726 & 292.067 & 292.268 & 292.468 & 0.02 & 0.99 \\
HO-CH$_3$ & O                     & 538.828 & 538.917 & 538.905 & 538.905$\dagger$      \\
F-CH$_3$ & C                      & 292.757 & 293.107 & 293.251 & 293.518 & 0.02 & 0.99 \\
F-CH$_3$ & F                      & 692.737 & 692.677 & 692.637 & 692.588 & 0.00 & 1.00 \\
F$_2$ & F                         & 696.959 & 696.801 & 696.706 & 696.588* & 0.01 & 1.00 \\
&                                 & 696.965 & 696.807 & 696.735 & 696.621 & 0.00 & 1.00 \\
H$_2$C=CH$_2$ & C                 & 290.306 & 290.601 & 290.701 & 290.872* & 0.01 & 1.00 \\
&                                 & 290.266 & 290.552 & 290.637 & 290.802 & 0.01 & 0.99 \\
H$_2$C=O & C                      & 293.772 & 294.130 & 294.275 & 294.498 & 0.00 & 1.00 \\
H$_2$C=O & O                      & 539.235 & 539.314 & 539.370 & 539.420 & 0.01 & 0.99 \\
HC$\equiv$CH & C                  & 290.695 & 290.942 & 291.116 & 291.274* & 0.02 & 0.99 \\
&                                 & 290.622 & 290.928 & 291.027 & 291.216 & 0.01 & 1.00 \\
H$_3$C-CH$_2$-CH$_3$ & -CH$_3$    & 289.903 & 290.226 & 290.305 & 290.487* & 0.02 & 0.99 \\
&                                 & 289.903 & 290.199 & 290.272 & 290.439 & 0.02 & 0.99 \\
H$_3$C-CH$_2$-CH$_3$ & -CH$_2$-   & 290.025 & 290.367 & 290.421 & 290.612 & 0.03 & 0.97 \\
H$_3$C-CH$_2$-NH$_2$ & H$_3$C-    & 289.940 & 290.272 & 290.357 & 290.546 & 0.02 & 0.99 \\
H$_3$C-CH$_2$-NH$_2$ & -CH$_2$-   & 290.788 & 291.118 & 291.305 & 291.496 & 0.01 & 1.00 \\
H$_3$C-CH$_2$-NH$_2$ & N          & 404.457 & 404.702 & 404.836 & 404.977 & 0.01 & 1.00 \\
H$_3$C-CH$_2$-OH & H$_3$C-        & 290.185 & 290.527 & 290.618 & 290.815 & 0.02 & 0.99 \\
H$_3$C-CH$_2$-OH & -CH$_2$-       & 291.580 & 291.925 & 292.065 & 292.266 & 0.00 & 1.00 \\
H$_3$C-CH$_2$-OH & O              & 538.472 & 538.568 & 538.646 & 538.703 & 0.01 & 0.97 \\
H$_3$C-CH$_2$-F & H$_3$C-         & 290.470 & 290.808 & 290.919 & 291.118 & 0.01 & 1.00 \\
H$_3$C-CH$_2$-F & -CH$_2$-        & 292.490 & 292.866 & 292.999 & 293.221 & 0.01 & 1.00 \\
H$_3$C-CH$_2$-F & F               & 692.194 & 692.151 & 692.124 & 692.098 & 0.00 & 0.99 \\
F$_2$CH$_2$ & C                   & 295.335 & 295.736 & 295.867 & 296.123 & 0.02 & 0.99 \\
F$_2$CH$_2$ & F                   & 693.499 & 693.458 & 693.433 & 693.406* & 0.00 & 0.99 \\
&                                 & 693.498 & 693.457 & 693.433 & 693.406 & 0.00 & 1.00 \\
H$_3$C-NH-CH$_3$ & N              & 404.501 & 404.699 & 404.671 & 404.671$\dagger$      \\
H$_3$C-NH-CH$_3$ & C              & 290.709 & 291.029 & 291.116 & 291.298 & 0.02 & 0.99 \\
&                                 & 290.709 & 291.033 & 291.132 & 291.317 & 0.01 & 0.99 \\
H$_3$C-O-CH$_3$ & O               & 538.389 & 538.451 & 538.437 & 538.437$\dagger$      \\
H$_3$C-O-CH$_3$ & C               & 291.534 & 291.870 & 292.010 & 292.206* & 0.00 & 1.00 \\
&                                 & 291.534 & 291.868 & 291.944 & 292.137 & 0.02 & 0.98 \\
H$_3$C-HC=CH$_2$ & H$_3$C-        & 290.215 & 290.518 & 290.543 & 290.716 & 0.03 & 0.95 \\
H$_3$C-HC=CH$_2$ & -HC=           & 290.157 & 290.451 & 290.557 & 290.728 & 0.01 & 1.00 \\
H$_3$C-HC=CH$_2$ & =CH$_2$        & 289.692 & 289.985 & 290.094 & 290.264 & 0.01 & 1.00 \\
H$_3$C-HC=O & H$_3$C-             & 290.786 & 291.104 & 291.295 & 291.491 & 0.01 & 0.99 \\
H$_3$C-HC=O & -HC                 & 293.202 & 293.556 & 293.726 & 293.941 & 0.00 & 1.00 \\
H$_3$C-HC=O & O                   & 538.249 & 538.341 & 538.478 & 538.539 & 0.03 & 0.89 \\
H$_3$C-C$\equiv$N & H$_3$C-       & 292.045 & 292.388 & 292.491 & 292.703 & 0.01 & 0.99 \\
H$_3$C-C$\equiv$N & -C$\equiv$    & 291.927 & 292.167 & 292.288 & 292.438 & 0.00 & 1.00 \\
H$_3$C-C$\equiv$N & N             & 404.988 & 405.238 & 405.399 & 405.558 & 0.01 & 0.99 \\
H$_2$N-HC=NH & H$_2$N-            & 405.610 & 405.821 & 406.064 & 406.199 & 0.04 & 0.93 \\
H$_2$N-HC=NH & C                  & 292.153 & 292.520 & 292.668 & 292.890 & 0.00 & 1.00 \\
H$_2$N-HC=NH & =NH                & 403.631 & 403.870 & 403.956 & 404.100 & 0.01 & 1.00 \\
H$_2$N-HC=O & N                   & 405.859 & 406.262 & 406.325 & 406.569 & 0.03 & 0.97 \\
H$_2$N-HC=O & C                   & 293.389 & 293.750 & 293.808 & 294.028 & 0.03 & 0.97 \\
H$_2$N-HC=O & O                   & 537.223 & 537.336 & 537.447 & 537.521 & 0.02 & 0.96 \\
HO-HC=O & HO-                     & 540.369 & 540.477 & 540.539 & 540.611 & 0.00 & 1.00 \\
HO-HC=O & C                       & 294.668 & 295.039 & 295.194 & 295.433 & 0.00 & 1.00 \\
HO-HC=O & =O                      & 538.471 & 538.577 & 538.701 & 538.774 & 0.02 & 0.93 \\
F-HC=CH$_2$ & F                   & 693.045 & 693.015 & 692.998 & 692.979 & 0.00 & 1.00 \\
F-HC=CH$_2$ & -HC=                & 292.686 & 293.044 & 293.126 & 293.345 & 0.02 & 0.98 \\
F-HC=CH$_2$ & =CH$_2$             & 290.306 & 290.627 & 290.664 & 290.859 & 0.03 & 0.96 \\
H$_2$N-N=CH$_2$ & H$_2$N-         & 405.472 & 406.000 & 406.121 & 406.437 & 0.03 & 0.98 \\
H$_2$N-N=CH$_2$ & -N=             & 405.746 & 405.985 & 406.089 & 406.233 & 0.00 & 1.00 \\
H$_2$N-N=CH$_2$ & C               & 290.541 & 290.863 & 291.018 & 291.214 & 0.00 & 1.00 \\
HO-N=CH$_2$ & O                   & 539.580 & 539.679 & 539.724 & 539.786 & 0.00 & 1.00 \\
HO-N=CH$_2$ & N                   & 406.674 & 406.921 & 407.092 & 407.250 & 0.02 & 0.99 \\
HO-N=CH$_2$ & C                   & 291.138 & 291.468 & 291.628 & 291.835 & 0.00 & 1.00 \\
H$_3$C-C$\equiv$CH & H$_3$C       & 291.061 & 291.396 & 291.502 & 291.703 & 0.01 & 0.99 \\
H$_3$C-C$\equiv$CH & -C$\equiv$   & 290.287 & 290.502 & 290.681 & 290.815 & 0.02 & 0.97 \\
H$_3$C-C$\equiv$CH & $\equiv$CH   & 289.727 & 289.968 & 289.943 & 290.083 & 0.04 & 0.85 \\
-CH$_2$-CH$_2$-CH$_2$- & C        & 290.117 & 290.419 & 290.552 & 290.728* & 0.00 & 1.00 \\
&                                 & 290.117 & 290.419 & 290.552 & 290.728 & 0.00 & 1.00 \\
&                                 & 290.142 & 290.402 & 290.568 & 290.721 & 0.02 & 0.99 \\
-CH$_2$-CH$_2$-NH- & C            & 291.001 & 291.301 & 291.433 & 291.611* & 0.00 & 1.00 \\
&                                 & 290.983 & 291.274 & 291.446 & 291.620 & 0.01 & 0.99 \\
-CH$_2$-CH$_2$-NH- & N            & 404.971 & 405.136 & 405.180 & 405.277 & 0.01 & 0.99 \\
-CH$_2$-CH$_2$-O- & C             & 291.873 & 292.223 & 292.276 & 292.484* & 0.03 & 0.97 \\
&                                 & 291.854 & 292.178 & 292.351 & 292.549 & 0.01 & 1.00 \\
-CH$_2$-CH$_2$-O- & O             & 538.783 & 538.853 & 538.989 & 539.036 & 0.03 & 0.85 \\
\hline
\label{tab:GW_32}\\
\end{longtable}  

\begin{table}[]
    \centering
    \begin{tabular}{c c c c c c}
        \hline                    
        Quantity                      & C          & N          & O          & F         & all  \\
        \hline
        Non-converged molecules ($\%$) & 201 (1.56) & 118 (1.39) & 146 (1.88) & 26 (0.80) & 201 (1.56)   \\
        Non-converged atoms ($\%$)     & 116 (0.20) & 28 (0.18)  & 65 (0.60)  &  9 (0.22)    &  218 (0.25) \\
        Distribution ($sp^3$,$sp^2$,$sp$) &(29,84,3)&(2,25,1)&(2,63,-)&(9,-,-)&(42,172,4) \\
        \hline                        
    \end{tabular}
    \caption{Summary of non-converged systems in the bigQM7$\omega$ dataset. Cations with a 1$s$-hole that fail to converge during the SCF procedure are labeled as 'Non-converged atoms.' Molecules containing one or more non-converged atoms are denoted 'Non-converged molecules.' The percentage relative to the total number of atoms/molecules in the bigQM7$\omega$ dataset is provided in parentheses. A breakdown of the atoms by hybridization is also included.
    } 
    \label{tab:201_summary}
\end{table}

\centering
\begin{longtable}{c c c}

     \caption{Details of 201 molecules exhibiting failure in SCF convergence for core-ionized cations. The column Molecule id denotes the title of the molecule in the XYZ format collected in the bigQM7$\omega$ dataset. SMILES representation and index of atoms for which core-ionized cations exhibiting convergence failure are also provided.}
     \label{tab:201_smiles} \\
     \hline
     Molecule id & SMILES & Atomic index \\
     \hline
     bigQM7$\omega\_$000312 &  CNCC=O       &  5        \\        
     bigQM7$\omega\_$000339 &  C=CNC=N         &  4        \\    
     bigQM7$\omega\_$000604 &  CNCC(C)=O       &  4        \\
     bigQM7$\omega\_$000635 &  CC(=O)CC=O      &  5        \\
     bigQM7$\omega\_$000649 &  CC(=N)NC=C      &  2        \\
     bigQM7$\omega\_$000670 &  NC(=N)ON=C      &  5        \\
     bigQM7$\omega\_$000836 &  CC(=C)C(C)=C    &  1,5      \\
     bigQM7$\omega\_$000841 &  CC(=C)C(F)=C    &  5        \\
     bigQM7$\omega\_$001073 &  ON(C=C)C=N      &  5        \\
     bigQM7$\omega\_$001162 &  NCC(=C)C=C      &  1        \\
     bigQM7$\omega\_$001163 &  OCC(=C)C=C      &  2        \\
     bigQM7$\omega\_$001166 &  FCC(=C)C=C      &  2        \\
     bigQM7$\omega\_$001173 &  OCC(=O)C=O      &  6        \\
     bigQM7$\omega\_$001193 &  COC(=C)C=O      &  6        \\
     bigQM7$\omega\_$001220 &  C=CC(=O)C=O     &  5,6      \\
     bigQM7$\omega\_$001485 &  O=c1nn[nH]o1    &  4        \\
     bigQM7$\omega\_$001736 &  CCNCC=O         &  6        \\
     bigQM7$\omega\_$001754 &  C=CCCC=O        &  6        \\
     bigQM7$\omega\_$001756 &  O=CCCC=O        &  1,2,5,6  \\
     bigQM7$\omega\_$001775 &  C=NOCC=O        &  2        \\
     bigQM7$\omega\_$001838 &  ON=CCC=O        &  5        \\
     bigQM7$\omega\_$001867 &  ON=CON=C        &  5        \\
     bigQM7$\omega\_$001962 &  O=CC=CC=O       &  5        \\
     bigQM7$\omega\_$002038 &  O=CC1CNC1       &  2        \\
     bigQM7$\omega\_$002041 &  O=CC1COC1       &  2        \\
     bigQM7$\omega\_$002540 &  CN(C)CC(O)=O    &  5        \\
     bigQM7$\omega\_$002552 &  CC(=O)CC(C)=O   &  2,5      \\
     bigQM7$\omega\_$002554 &  CC(=O)CC(N)=O   &  2        \\
     bigQM7$\omega\_$002765 &  CC(F)(CO)C=O    &  6        \\
     bigQM7$\omega\_$002830 &  FC(F)(C=O)C=O   &  7        \\
     bigQM7$\omega\_$002945 &  CC(N)CON=C      &  7        \\
     bigQM7$\omega\_$003062 &  CC(=C)CCC=O     &  7        \\
     bigQM7$\omega\_$003065 &  CC(=O)CCC=O     &  7        \\
     bigQM7$\omega\_$003077 &  FC(=C)CCC=O     &  7        \\
     bigQM7$\omega\_$003094 &  NC(=O)CNN=C     &  6        \\
     bigQM7$\omega\_$003099 &  CC(=O)COC=C     &  2        \\
     bigQM7$\omega\_$003112 &  NC(=O)CON=C     &  7        \\
     bigQM7$\omega\_$003125 &  NC(=O)NCC=O     &  7        \\
     bigQM7$\omega\_$003136 &  CC(=O)NOC=O     &  6        \\
     bigQM7$\omega\_$003416 &  CC(=C)C=CC=O    &  1        \\
     bigQM7$\omega\_$003419 &  CC(=O)C=CC=O    &  6        \\
     bigQM7$\omega\_$003422 &  NC(=N)C=CC=O    &  7        \\
     bigQM7$\omega\_$003431 &  FC(=C)C=CC=O    &  1        \\
     bigQM7$\omega\_$003432 &  FC(=C)C=CC$\#$N    &  1        \\
     bigQM7$\omega\_$003521 &  FC(=C)C$\#$CC=C    &  1        \\
     bigQM7$\omega\_$003990 &  CC(CN)ON=C      &  6        \\
     bigQM7$\omega\_$004072 &  COCC(F)C=O      &  6        \\
     bigQM7$\omega\_$004115 &  CC(CC=O)C=O     &  4        \\
     bigQM7$\omega\_$004123 &  OC(CC=C)C=O     &  7        \\
     bigQM7$\omega\_$004125 &  OC(CC=O)C=O     &  5,7      \\
     bigQM7$\omega\_$004132 &  FC(CC=C)C=O     &  7        \\
     bigQM7$\omega\_$004142 &  CC(NC=O)C=O     &  7        \\
     bigQM7$\omega\_$004160 &  CN(CC=O)C=O     &  5        \\
     bigQM7$\omega\_$004187 &  FC(CC$\#$C)C=O     &  7        \\
     bigQM7$\omega\_$004437 &  CCC(=N)NC=C     &  3        \\
     bigQM7$\omega\_$004446 &  FCC(=N)NC=C     &  3        \\
     bigQM7$\omega\_$004522 &  CNC(=N)ON=C     &  7        \\
     bigQM7$\omega\_$004523 &  NNC(=N)ON=C     &  6        \\
     bigQM7$\omega\_$004564 &  CNCC(=C)C=C     &  2        \\
     bigQM7$\omega\_$004565 &  CNCC(=C)C=O     &  7        \\
     bigQM7$\omega\_$004569 &  CNCC(=O)C$\#$N     &  7        \\
     bigQM7$\omega\_$004570 &  COCC(=C)C=C     &  3        \\
     bigQM7$\omega\_$004580 &  CCNC(=O)C=O     &  7        \\
     bigQM7$\omega\_$004589 &  CON=C(N)C=O     &  7        \\
     bigQM7$\omega\_$004595 &  CCOC(=C)C=O     &  7        \\
     bigQM7$\omega\_$004602 &  C=CC(=C)CC$\#$N    &  6        \\
     bigQM7$\omega\_$004605 &  C=C(CC=O)C=O    &  5        \\
     bigQM7$\omega\_$004614 &  O=CCC(=O)C=O    &  5        \\
     bigQM7$\omega\_$004618 &  C=CC(=C)NC=N    &  7        \\
     bigQM7$\omega\_$004619 &  C=CC(=C)NC=O    &  7        \\
     bigQM7$\omega\_$004627 &  C=CNC(=N)C=O    &  6        \\
     bigQM7$\omega\_$004631 &  N=C(NC=O)C=O    &  6        \\
     bigQM7$\omega\_$004674 &  CNCC(=O)C$\#$C     &  4        \\
     bigQM7$\omega\_$004693 &  C=CNC(=N)C$\#$C    &  4        \\
     bigQM7$\omega\_$004842 &  CN=CN(N)C=C     &  3        \\
     bigQM7$\omega\_$004880 &  CCC(=C)C=CC     &  1        \\
     bigQM7$\omega\_$004948 &  CC=CC(=C)C=O    &  7        \\
     bigQM7$\omega\_$004959 &  FC=CC(=O)C=O    &  7        \\
     bigQM7$\omega\_$005027 &  CC$\#$CC(=O)C=O    &  7        \\
     bigQM7$\omega\_$005089 &  FC=C(F)CC=O     &  7        \\
     bigQM7$\omega\_$005093 &  CC(CC=O)=NO     &  4        \\
     bigQM7$\omega\_$005120 &  CC(ON=C)=NO     &  4        \\
     bigQM7$\omega\_$005271 &  CC(C=C)=CCO     &  1        \\
     bigQM7$\omega\_$005272 &  CC(C=C)=CCF     &  1        \\
     bigQM7$\omega\_$005293 &  COC=C(C)C=C     &  1        \\
     bigQM7$\omega\_$005307 &  CC(C=C)=CC$\#$N    &  1        \\
     bigQM7$\omega\_$005315 &  NC(C=C)=CC$\#$N    &  3        \\
     bigQM7$\omega\_$005317 &  NC(=CC=O)C$\#$N    &  6        \\
     bigQM7$\omega\_$005668 &  CC(C=O)C(F)=C   &  3        \\
     bigQM7$\omega\_$005705 &  FC(C=O)C(F)=C   &  4        \\
     bigQM7$\omega\_$005709 &  CN(C=C)C(C)=N   &  5        \\
     bigQM7$\omega\_$005723 &  CC(=N)N(N)C=C   &  2        \\
     bigQM7$\omega\_$005896 &  CC(O)C(=C)C=C   &  2,3      \\
     bigQM7$\omega\_$005965 &  CCC(=C)C(F)=C   &  6        \\
     bigQM7$\omega\_$005982 &  OCC(=O)C(O)=O   &  5        \\
     bigQM7$\omega\_$005991 &  CC(=O)C(=N)NO   &  2        \\
     bigQM7$\omega\_$006035 &  CC(=C)C(=C)C$\#$N  &  1        \\
     bigQM7$\omega\_$006052 &  CC(=C)C(=O)C=O  &  7        \\
     bigQM7$\omega\_$006058 &  NC(=N)C(=O)C=O  &  5,7      \\
     bigQM7$\omega\_$006064 &  OC(=O)C(=O)C=O  &  5        \\
     bigQM7$\omega\_$006294 &  CC1(CNC1)C=O    &  6        \\
     bigQM7$\omega\_$006304 &  CC1(COC1)C=O    &  6        \\
     bigQM7$\omega\_$006311 &  FC1(COC1)C=O    &  6        \\
     bigQM7$\omega\_$006586 &  O=C1CNC(=O)C1   &  2        \\
     bigQM7$\omega\_$006653 &  CC1CC(=C)C=C1   &  1        \\
     bigQM7$\omega\_$006671 &  CN1CC(=C)C=N1   &  1        \\
     bigQM7$\omega\_$006704 &  C=C1NC(=N)C=C1  &  4,7      \\
     bigQM7$\omega\_$006709 &  C=C1NC(=N)N=C1  &  6        \\
     bigQM7$\omega\_$006715 &  C=C1OC(=C)C=C1  &  7        \\
     bigQM7$\omega\_$006956 &  CC1=CC(N)C=C1   &  1        \\
     bigQM7$\omega\_$006959 &  OC1C=CC(F)=C1   &  6        \\
     bigQM7$\omega\_$006961 &  FC1C=CC(F)=C1   &  6        \\
     bigQM7$\omega\_$007033 &  CC1=CC(=O)C=N1  &  5        \\
     bigQM7$\omega\_$007036 &  NC1=CC(=N)N=C1  &  6        \\
     bigQM7$\omega\_$007476 &  O=C1COC=CC1     &  2        \\
     bigQM7$\omega\_$007576 &  C=C1CCCC=C1     &  3        \\
     bigQM7$\omega\_$007587 &  C=C1COCC=C1     &  3        \\
     bigQM7$\omega\_$007601 &  CC1CC=CC=C1     &  1        \\
     bigQM7$\omega\_$007637 &  Nc1cncnn1       &  4        \\
     bigQM7$\omega\_$007709 &  O=c1cn[nH]nn1   &  7        \\
     bigQM7$\omega\_$007961 &  O=C1CCCC1=O     &  1,7      \\
     bigQM7$\omega\_$007970 &  N=C1NCCC1=O     &  7        \\
     bigQM7$\omega\_$007980 &  C=C1CCOC1=C     &  5        \\
     bigQM7$\omega\_$008035 &  FC1CC=CC1=C     &  1        \\
     bigQM7$\omega\_$008060 &  Nn1[nH]ncc1=N   &  5        \\
     bigQM7$\omega\_$008063 &  On1[nH]ncc1=O   &  5        \\
     bigQM7$\omega\_$008071 &  Cn1[nH]nnc1=O   &  5        \\
     bigQM7$\omega\_$008073 &  Nn1[nH]nnc1=O   &  5        \\
     bigQM7$\omega\_$008082 &  Cn1oncc1=N      &  4        \\
     bigQM7$\omega\_$008097 &  Nn1onnc1=O      &  4        \\
     bigQM7$\omega\_$008195 &  C=C1CC=CC1=C    &  5        \\
     bigQM7$\omega\_$008202 &  O=C1NC=CC1=O    &  2        \\
     bigQM7$\omega\_$008206 &  O=C1NN=CC1=O    &  2        \\
     bigQM7$\omega\_$008207 &  C=C1N=CNC1=N    &  6        \\
     bigQM7$\omega\_$008209 &  N=C1NC=NC1=N    &  4        \\
     bigQM7$\omega\_$008212 &  C=C1OC=CC1=C    &  2,4,6    \\
     bigQM7$\omega\_$008254 &  FC1=CCCC1=C     &  1        \\
     bigQM7$\omega\_$008403 &  CC1=NC=CC1=O    &  7        \\
     bigQM7$\omega\_$008759 &  O=CNC1CNC1      &  2        \\
     bigQM7$\omega\_$008760 &  C=NNC1CNC1      &  2        \\
     bigQM7$\omega\_$008898 &  NC1CC1OC=O      &  6        \\
     bigQM7$\omega\_$008908 &  CC1NC1CC=O      &  6,7      \\
     bigQM7$\omega\_$009003 &  CC(=O)C1CNC1    &  2        \\
     bigQM7$\omega\_$009005 &  NC(=O)C1CNC1    &  2        \\
     bigQM7$\omega\_$009367 &  CNC(C=O)C$\#$N     &  5        \\
     bigQM7$\omega\_$009421 &  O=CC(C=O)C$\#$N    &  5        \\
     bigQM7$\omega\_$009441 &  O=CC(C=O)C$\#$C    &  5        \\
     bigQM7$\omega\_$009591 &  CNC(C=O)=NC     &  5        \\
     bigQM7$\omega\_$009831 &  CCOCCC=O        &  7        \\
     bigQM7$\omega\_$009837 &  CCCNCC=O        &  7        \\
     bigQM7$\omega\_$009844 &  FCCNCC=O        &  6        \\
     bigQM7$\omega\_$009895 &  C=CCCCC=O       &  7        \\
     bigQM7$\omega\_$009902 &  N=CNCCC=O       &  7        \\
     bigQM7$\omega\_$009904 &  O=CCCNC=O       &  1        \\
     bigQM7$\omega\_$009915 &  C=CCCON=C       &  6        \\
     bigQM7$\omega\_$009916 &  C=NOCCC=O       &  1,2      \\
     bigQM7$\omega\_$009921 &  C=COCON=C       &  7        \\
     bigQM7$\omega\_$009963 &  O=CCOCC$\#$C       &  1        \\
     bigQM7$\omega\_$010050 &  FC=CCCC=O       &  7        \\
     bigQM7$\omega\_$010054 &  ON=CCCC=O       &  7        \\
     bigQM7$\omega\_$010081 &  NN=CNCC=C       &  7        \\
     bigQM7$\omega\_$010105 &  NN=COCC=C       &  6        \\
     bigQM7$\omega\_$010110 &  CC=NOCC=O       &  7        \\
     bigQM7$\omega\_$010265 &  CNCC=CC=O       &  7        \\
     bigQM7$\omega\_$010267 &  COCC=CC=C       &  1        \\
     bigQM7$\omega\_$010268 &  COCC=CC=O       &  6        \\
     bigQM7$\omega\_$010274 &  CNCC=NN=C       &  7        \\
     bigQM7$\omega\_$010312 &  OCC=CCC=O       &  7        \\
     bigQM7$\omega\_$010327 &  OCC=CNC=N       &  6        \\
     bigQM7$\omega\_$010328 &  FCC=CNC=N       &  6        \\
     bigQM7$\omega\_$010374 &  O=CCC=CC=O      &  6        \\
     bigQM7$\omega\_$010406 &  C=NOC=NN=C      &  2        \\
     bigQM7$\omega\_$010474 &  CC=CC=CC=C      &  1        \\
     bigQM7$\omega\_$010546 &  COCC$\#$CC=C       &  1        \\
     bigQM7$\omega\_$010570 &  O=CCC$\#$CC=O      &  6        \\
     bigQM7$\omega\_$010575 &  O=CCC$\#$CC$\#$C      &  1        \\
     bigQM7$\omega\_$010892 &  NCC1CN1N=C      &  6        \\
     bigQM7$\omega\_$011104 &  O=CC1CC=NN1     &  1        \\
     bigQM7$\omega\_$011113 &  O=CC1CC=NO1     &  2        \\
     bigQM7$\omega\_$011115 &  C=CC1OC=CO1     &  2        \\
     bigQM7$\omega\_$011178 &  ON=c1onno1      &  5        \\
     bigQM7$\omega\_$011223 &  O=CC1COC=C1     &  1        \\
     bigQM7$\omega\_$011229 &  O=CC1NCC=C1     &  2        \\
     bigQM7$\omega\_$011236 &  C=CN1CCN=C1     &  7        \\
     bigQM7$\omega\_$011349 &  N=Cn1cccc1      &  2        \\
     bigQM7$\omega\_$011351 &  C=Nn1cccc1      &  2        \\
     bigQM7$\omega\_$011389 &  O=Nc1cc[nH]c1   &  1        \\
     bigQM7$\omega\_$011461 &  C=CC1=CCCC1     &  7        \\
     bigQM7$\omega\_$011469 &  C=CC1=CCNC1     &  7        \\
     bigQM7$\omega\_$011472 &  C=CC1=CCOC1     &  7        \\
     bigQM7$\omega\_$011483 &  N$\#$CC1=NCCN1     &  2        \\
     bigQM7$\omega\_$011623 &  O=Cc1ncc[nH]1   &  2        \\
     bigQM7$\omega\_$011638 &  O=Cc1ccno1      &  2        \\
     bigQM7$\omega\_$011672 &  CCC1=CCC=C1     &  1        \\
     bigQM7$\omega\_$011742 &  NCc1nnon1       &  5        \\
     bigQM7$\omega\_$011879 &  OC(C=O)C1CN1    &  3        \\
     bigQM7$\omega\_$011985 &  C=CC(=C)C1CO1   &  5,6      \\
     bigQM7$\omega\_$011995 &  O=CC(=O)N1CC1   &  1        \\
     bigQM7$\omega\_$012467 &  N1C=CC=NC=N1    &  2,4      \\
     bigQM7$\omega\_$012468 &  N1C=NC=NC=N1    &  2        \\
     bigQM7$\omega\_$012799 &  NC1C2NC1C=C2    &  2        \\

     \hline

\end{longtable}

\bibliography{ref}